\documentclass{emulateapj}

\begin{document}

\title{Jet-Powered Molecular Hydrogen Emission from Radio Galaxies}

\author{Patrick Ogle$^1$, Francois Boulanger$^2$, Pierre Guillard$^1$, Daniel A. Evans$^3$, Robert Antonucci$^4$, P. N. Appleton$^5$, Nicole Nesvadba$^2$, 
        \& Christian Leipski$^6$}

\affil{$^1$Spitzer Science Center, California Institute of Technology, 
       Mail Code 220-6, Pasadena, CA 91125}

\affil{$^2$Institut d'Astrophysique Spatiale, Universite Paris Sud, Bat. 121, 91405 Orsay Cedex, France}

\affil{$^3$MIT Center for Space Research, Cambridge, MA 02139}

\affil{$^4$Physics Dept., University of California, Santa Barbara, CA 93106}

\affil{$^5$NASA Herschel Science Center, California Institute of Technology, 
       Mail Code 100-22, Pasadena, CA 91125}

\affil{$^6$Max-Planck Institut f\"ur Astronomie (MPIA), K\"onigstuhl 17, D-69117 
  Heidelberg, Germany}

\email{ogle@ipac.caltech.edu}

\shorttitle{Molecular Hydrogen in Radio Galaxies} 
\shortauthors{Ogle et al.}

\begin{abstract}

H$_2$ pure-rotational emission lines are detected from warm (100-1500 K) molecular gas in 17/55 (31\% of) radio 
galaxies at redshift $z<0.22$ observed with the {\it Spitzer} IR Spectrograph. The summed H$_2$ 0-0 S(0)-S(3) line 
luminosities are $L(\mathrm{H}_2)=7\times 10^{38}-2\times 10^{42}$ erg s$^{-1}$, yielding warm H$_2$ masses up to 
$2\times 10^{10} M_\odot$. These radio galaxies, of both FR radio morphological types, help to firmly 
establish the new class of radio-selected molecular hydrogen emission galaxies (radio MOHEGs). MOHEGs have extremely 
large H$_2$ to 7.7 $\mu$m PAH emission ratios: $L$(H$_2$)/$L(\mathrm{PAH7.7})=0.04-4$, up to a factor 300 greater than the 
median value for normal star-forming galaxies. In spite of large H$_2$ masses, MOHEGs appear to be 
inefficient at forming stars, perhaps because the molecular gas is kinematically unsettled and turbulent. Low-luminosity 
mid-IR continuum emission together with low-ionization emission line spectra indicate low-luminosity AGNs in all but 3 radio 
MOHEGs. The AGN X-ray emission measured with {\it Chandra} is not luminous enough to power the H$_2$ emission from MOHEGs. 
Nearly all radio MOHEGs belong to clusters or close pairs, including 4 cool core clusters (Perseus, Hydra, A 2052, and A 2199). 
We suggest that the H$_2$ in radio MOHEGs is delivered in galaxy collisions or cooling flows, then heated by radio jet feedback 
in the form of kinetic energy dissipation by shocks or cosmic rays.

\end{abstract}


\section{Introduction}

\subsection{AGN Radio Jet Feedback}

AGN feedback on the interstellar and intracluster media (ISM and ICM) is thought to be an important factor regulating gas accretion, 
star formation, and AGN activity in galaxies \citep{gzs04,so04,hhc06,csw06,co07,mh08}. This may occur through either a sub-relativistic wind 
in radio-quiet quasars or a relativistic radio jet in the case of radio-loud quasars and radio galaxies. In the latter case, the kinetic energy 
of the jet integrated over its lifetime is typically more than enough to unbind the ISM of even the most massive galaxy. While much effort
has been devoted to simulating the impact of jets on the hot ICM \citep[e.g.,][]{zmb05}, less is known about the jet-ISM interaction.
Recent simulations have shown that a radio jet may couple strongly to an inhomogeneous, clumpy ISM until the radio lobes break free into 
intergalactic space \citep{sb07}. In this process, the jet drives radiative shocks into dense clumps but does not ablate them, while less 
dense gas may be ablated and advected to larger radii. While such detailed simulations have been made of the impact of jets on the atomic 
and ionized components of the ISM, the impact on the molecular gas phase has not been considered in any detail. 

\cite{oaa07} found that the mid-IR spectrum of radio galaxy 3C 326 N displays high-luminosity H$_2$ emission, coupled with weak PAH 
emission indicative of a low star formation rate. We presented evidence that 3C 326 N belongs to an interacting galaxy pair and suggested 
that the copious warm H$_2$ is heated by accretion from the companion 3C 326 S. Further investigation revealed the presence of fast neutral 
gas outflow, possibly driven by the radio jet \citep{nbs10}. Therefore, a strong interaction between the radio jet and molecular gas, 
rather than accretion, may power the H$_2$ emission. 

H$_2$ emission may prove to be a key in understanding how radio jet feedback impacts the molecular gas phase and thereby 
suppresses star formation in massive galaxies. The most likely H$_2$ heating mechanism is jet-driven shocks in a multiphase interstellar medium. 
Cosmic ray heating and X-ray emission from the AGN are alternative heating mechanisms that we will consider.

\subsection{H$_2$ Emission from Galaxies}

Rotational and ro-vibrational H$_2$ transitions are important cooling channels for warm molecular gas, observed as 
a series of infrared emission lines. These transitions are highly forbidden because of the zero permanent
dipole moment of H$_2$. In normal star-forming galaxies with total H$_2$ to total (3-1100 $\mu$m) IR ratios 
$L$(H$_2$)/$L_\mathrm{TIR}<0.001$, these lines are thought to be emitted by stellar photodissociation 
regions \citep{rkl02,hah06,rhh07}. 

Much stronger molecular hydrogen emission galaxies (MOHEGs) with 
$L$(H$_2$)/$L_\mathrm{TIR}>0.001-0.1$ have been revealed by recent {\it Spitzer} observations \citep{oaa07}.  Extreme 
H$_2$ emission has been found in a variety of galaxy types so far, including luminous infrared galaxies \citep{lsg03}, 
colliding galaxies \citep{a06}, central dominant cluster galaxies \citep{e06,dm09}, and radio galaxies \citep{oaa07,wsd10}.  
Enhanced H$_2$ emission has been observed in AGNs with ISO \citep{rkl02}, and more recently from AGNs 
in the {\it Spitzer} SINGS survey \citep{rhh07}. Relatively strong H$_2$ emission is also seen in PG quasars with strong PAH emission
\citep{sls06,vrk09}.  AGN X-ray heating or shocks have been cited as possible ways to power the H$_2$ emission. 

The first galaxy discovered with strong ro-vibrational and rotational H$_2$ lines ($2\times 10^{42}$ erg s$^{-1}$) 
is the luminous infrared galaxy (LIRG) NGC 6240 \citep{jw84, lsg03, abs06}. Keck Adaptive Optics and HST NICMOS imaging in near-IR  
H$_2$ ro-vibrational lines reveal a complex network of hot molecular gas filaments embedded in a larger network of 
H$\alpha$ emission \citep{mcm05}. This galaxy has a binary active nucleus \citep{kbh03} and is thought to be a late stage 
major merger of two gas-rich spiral galaxies. Shock-heating driven by the merger may power the strong 
H$_2$ emission spectrum. 

~\cite{a06} find high luminosity H$_2$ emission from an intergalactic shock between colliding galaxies in Stephan's Quintet. The 
X-ray emission from the shock, though conspicuous, is too weak to power the H$_2$ lines. Instead, it is likely that the H$_2$  
emission is powered by kinetic energy dissipation through turbulence in a multi-phase post-shock layer \citep{gb09}. In this case 
the H$_2$ emission provides ten times the luminosity of the next brightest Mid-IR line [Si {\sc ii}], demonstrating that it
is likely the dominant coolant in this group-wide shock \citep{ca10}.

H$_2$ ro-vibrational emission from 1000-2000 K molecular gas is seen from the brightest central galaxies in some galaxy clusters \citep{d00}. 
Spitzer observations have shown these galaxies also display rotational lines from H$_2$ at lower (100-1500 K) temperatures \citep{e06,jhf07}. 
The warm molecular gas is often associated with extended optical emission line filaments. Extensive networks of such filaments are found in 
30-50\% of X-ray luminous, rich galaxy clusters, associated with the central-dominant (cD) galaxy \citep{l70,h81,chj83,jfn87,hbv89,cae99}. 

The origin, confinement, and excitation mechanism for cluster emission line filaments are longstanding mysteries \citep{m57,bb65}. 
The filaments in Perseus A may be formed in-situ by the interaction of the radio lobes with the hot intracluster medium \citep{cgw01}. 
Since the filaments are dusty and have near-solar metal abundances, it is more likely that they are composed of cool gas dredged 
up by the radio jet from the center of the galaxy \citep{cfj00,ffh09}. Galaxy collisions and ram-pressure stripping are two other
ways to create intracluster filaments of atomic and molecular gas \citep{ktc08,srr10}. 

The gas filaments surrounding Per A are $\sim 100$ times more luminous than expected from a 
cooling flow and therefore must be powered externally \citep{chj83}. The H$\alpha$ filaments consist of extremely narrow 
threads, most likely confined by a magnetic field \citep{fjs08}. Kinetic energy dissipation by magnetohydrodynamic 
(MHD) shocks or dissipative MHD waves driven by the radio lobes may power the atomic and molecular emission line spectra 
\citep{ks79,hbv89,hcj05,gb09}.  Molecular gas heating by cosmic rays has also been suggested \citep{ffh08,ffh09}, 
and may help to explain some anomalous atomic forbidden emission line ratios. 

\subsection{Overview}

We present a large {\it Spitzer} spectroscopic survey of 55 radio galaxies. We find a large number of H$_2$ 
luminous galaxies (17), which belong to and further define the MOHEG class. We use mid-IR and optical forbidden lines, aromatic 
features, and 24 $\mu$m continuum emission to investigate the gas and dust content and H$_2$ power source in these radio-selected MOHEGs. 
We describe the radio galaxy sample, observations, and spectra in Sections 2-4. We estimate warm H$_2$ masses using excitation diagrams in 
Section 5. We explore the dust emission properties of the sample in Sections 6 and 7. We discuss the H$_2$/PAH diagnostic ratio and use this 
to redefine the MOHEG class in Sections 8 and 9. We estimate star formation rates and use ionic forbidden line ratios to diagnose AGN 
activity in Sections 10 and 11. We use {\it Chandra} observations to rule out X-ray heating by the AGN in Section 12. The host galaxy
morphologies and environments are discussed in Section 13. Radio jet morphologies are considered in Section 14. Finally, we discuss evidence in
support of radio jet feedback in Section 15.1, cosmic ray heating in Section 15.2, and the origin of the molecular gas in Section 15.3.

\section{Sample}

We conducted a survey of \cite{fr74} class-II radio galaxies and quasars from the 3CRR catalog with the {\it Spitzer} 
IRS \citep{owa06} and a similar survey of FR I radio galaxies \citep{loa09}. The FR II galaxies were selected from the 3CRR catalog
to form a redshift and flux-limited sample with $z<1$ and $S_\nu$(178 MHz)$>16.4$ Jy. The FR I galaxies form a representative (yet incomplete)
sample at $z<0.13$, selected from the 3CR catalog with a flux limit of  $S_\nu$(178 MHz)$>15$ Jy. 

From the above two samples, we selected all 23 FR I galaxies and the subsample of 19 FR II galaxies at $z<0.22$.  We excluded the blazars 
BL Lac and 3C 371, which have mid-IR spectra completely dominated by synchrotron emission, and the radio-intermediate quasar E1821+643.  
We retrieved the data for 8 additional FR II and 5 FR I radio galaxies in the {\it Spitzer} archive, for a total of 27 FR II and 28 FR I galaxies 
at $z<0.22$. The redshift limit of $z<0.22$ ensures that the full H$_2$ S(0)-S(7) line series falls in the {\it Spitzer} IRS wave band.

The radio galaxy sample is presented in Tables 1 and 2. Alternate names,
host galaxy properties, cluster membership, and interaction signatures are listed in Table 3 for H$_2$ detected radio galaxies. Throughout this
paper we assume a cosmology with Hubble constant $H_0=70$ km s$^{-1}$ Mpc$^{-1}$, matter density parameter $\Omega=0.3$, and dark 
energy density $\Omega_\Lambda=0.7$.

\section{Observations}

\subsection{Spitzer}

The radio galaxies in Tables 1 and 2 were observed with the Spitzer IR Spectrograph \citep[IRS;][]{h04}. 
Spectra were taken with the low-resolution ($R \sim 60-120$) modules Short-Low 2 (SL2), Short-Low 1 (SL1), Long-Low 2 (LL2), 
and Long-Low 1 (LL1) over the wavelength range 5-36.5 $\mu$m. At the galaxy redshifts of 0.002-0.22, 1\farcs0 corresponds to 
0.04-3.6 kpc. The slit widths are 3\farcs6, 3\farcs7, 10\farcs5, and 10\farcs7, respectively, corresponding to 0.15-13 kpc for 
SL and 0.42-38 kpc for LL.  The telescope was nodded to place the nucleus of each galaxy at the 1/3 and 2/3 positions 
along each slit. 

We began our analysis with the S15-S17 pipeline processed, basic calibrated data sets (BCDs). Off-slit observations were 
subtracted to remove background light. Spectra were extracted (using SPICE 2.1) within tapered regions matching the {\it Spitzer} 
point-spread function and including the first Airy ring (SL2: 7\farcs2 at 6 $\mu$m, SL1: 14\farcs4 at 12 $\mu$m,
LL2: 21\farcs7 at 16 $\mu$m, LL1: 36\farcs6 at 27 $\mu$m). We tried both regular and optimal extraction within these regions
\citep{nol07}, and selected the extraction with higher signal-to-noise for further analysis. For unresolved sources, the continuum 
and emission line fluxes from regular and optimal extractions are in agreement. 

The flux uncertainties produced by the {\it Spitzer} 
pipeline are known to be too large for faint sources. The pipeline (versions S18.7 and earlier) does not properly adjust the uncertainties 
for 'droop', a systematic, additive effect which is proportional to the total photocurrent incident upon the detector. We divided flux 
uncertainties for each spectral order by a constant factor of 0.3-2.1, to better reflect the observed noise characteristics. The uncertainty scale 
factor for each order was determined from the ratio of the mean pipeline uncertainties to the observed noise in emission line-free regions.

Spectral orders with low flux were adjusted upwards by factors of 1.0-2.0 to correct for interorder slit loss differences. Such differences 
may arise from pointing inaccuracies, even for unresolved sources. The differences in slit widths and wavelength-dependent extraction apertures 
may also lead to mismatches in continuum or line flux between orders for spatially resolved sources. Our spectral order matching will properly 
correct for this effect in the case that the extended continuum and line emission are cospatial and have similar spectra, such as might be the
case for extended star-forming regions. However, the emission line equivalent widths and line flux ratios between orders could be off by as much
as the interorder correction factor ($<2.0$) in the case of extended nebular or H$_2$ emission. This should not be an issue for any galaxies 
except Per A, 3C 293, and Cen A, the only ones where the emission lines and PAH features appear to be spatially extended along the spectrograph slits.
For all other sources, the mid-IR (MIR) emission (including the H$_2$ emission) is unresolved and resides within the central 2-40 kpc of the host 
galaxies.

We measured the emission line and PAH features of all galaxies in a systematic way using the spectral 
fitting tool PAHFIT \citep{sdd07}.  The continuum is modeled with a sum of starlight and thermal 
(blackbody) emission components, absorbed by a uniform dust screen. This is intended only as an empirical description of the 
continuum, not a physical model.  The spectrally resolved PAH 
features are fit by Drude profiles. The unresolved H$_2$ and forbidden atomic emission lines are fit by Gaussian profiles. 
Emission line and PAH fluxes from the galaxies are presented in Tables 4-8. We also measured narrow-band 24 $\mu$m  
fluxes (3.7$\sigma$-clipped average over 22.5-25.5 $\mu$m rest; Tables 1-2) from the IRS spectra and used them to estimate mid-IR continuum luminosities 
(Table 9). 

In principle, the continuum for each source should be fit with a dust radiative transfer model, but this is outside the scope of the 
present work. Instead, we tailored the continuum model in two cases where the simple multi-temperature black body plus 
screen extinction model did not work. For Perseus A, we added silicate emission features to the continuum. We introduced the capability 
to fit optically thin silicate emission features with PAHFIT, computing the silicate emission as blackbody emission multiplied by an 
empirical dust extinction curve. In the Cygnus A spectrum, 18 $\mu$m silicate absorption is unusually weak or absent relative to the 9 
$\mu$m silicate absorption, perhaps because of radiative transfer effects. Therefore we removed the 18 $\mu$m silicate absorption feature 
from the \cite{ct06} Galactic center extinction curve in order to model the absorption.

\subsection{Chandra}

In order to test the possibility of H$_2$ heating by X-rays from AGN, we compiled unabsorbed 2-10 keV 
nuclear X-ray luminosities measured by {\it Chandra} (Table 9). For several sources without published 
X-ray luminosities, we used archival {\it Chandra} data to measure them ourselves (Table 10). 
We reprocessed the {\it Chandra} data using {\sc CIAO} v4.1.2 with the CALDB v4.1.2 to create a
new level-2 events file filtered for the grades 0, 2, 3, 4, and 6 and with the 0.5$''$ pixel
randomization removed. To check for periods of high background, we extracted light curves for the
chip on which the source was placed, excluding point sources. Few periods of high background were
found in any observation, and so we used almost all available data in our analysis.

We extracted the nuclear spectra for each source from a small circle (typically of radius 5
pixels or 2.5$''$), and sampled background from an adjacent region free from point sources. In most cases,
the point-like emission from the nucleus was easily separated from any extended emission components, though
there may be significant contamination from the surrounding hot cluster gas in the case of the Zw3146 BCG. 
We do not consider the possibility of H$_2$ heating by X-ray emission from the extended intracluster medium or
X-ray emission from the extended radio source.

We adopted the standard X-ray spectral models for radio-loud AGN described by Evans et al. (2006) to fit
each spectrum. In the case of narrow-line radio galaxies (NLRGs), the canonical model takes the
form of the combination of a heavily absorbed power law and a soft, unabsorbed power law that \cite{ewh06} 
interpreted to originate at the base of an unresolved jet.  In Table 10 we present an observation log and 
summarize the best-fitting models for each source.

\section{H$_2$ Survey Results and Spectra}

We detect two or more H$_2$ emission lines from 17/55 (31\%) of the radio galaxies in our sample (Table 4). 
We require $>2.5\sigma$ detection of 2 lines at the systemic redshift.   A typical $2 \sigma$ H$_2$ S(1) detection limit for the sample is 
$8\times 10^{-16}$ erg s$^{-1}$ cm$^{-2}$  ($8\times 10^{-19}$ W m$^{-2}$), but this may vary with redshift and H$_2$ line contrast relative 
to the continuum. 

 H$_2$ line and PAH fluxes for galaxies with only one or no H$_2$ lines detected are presented in Table 6. Upper limits to their H$_2$ 
luminosities are estimated as $L_\mathrm{H2}<1.6 L$[H$_2$ 0-0 S(1)+ S(3)], where the correction factor 1.6 accounts for the S(0) + S(2) line 
flux and is the mean value of [H$_2$ S(0) + S(1) + S(2) + S(3)]/[H$_2$ S(1) + S(3)] for galaxies where all 4 lines were detected. We use this 
rather than the sum of 4 upper limits for S(1) through S(4), which would be an overestimate.

The {\it Spitzer} IRS spectra of the H$_2$ detections are presented in Figures 1-7, along with archival HST optical images for
most sources. Some of these spectra were previously presented by \cite{owa06}, \cite{oaa07}, and \cite{loa09}, but we show them here as well for 
completeness. The Cen A spectrum was originally presented by \cite{whh05}. H$_2$ pure-rotational lines, forbidden lines of O, Ne, Si, S, Ar, Cl, 
and Fe, and polycyclic aromatic hydrocarbon (PAH) features are seen in many of the spectra (Tables 4-8).  Galaxies with large 7.7 $\mu$m PAH 
equivalent width are presented in Figure 1 and those with small 7.7 $\mu$m PAH equivalent width in Figures 2-3. The mid-IR luminous radio galaxies 
Per A, Cyg A, 3C 433, and 3C 436, which contain dust-obscured quasars, are presented in Figure 4. The H$_2$ emission is difficult to see on top 
of the continuum emission from the galactic nuclei in some spectra. However, subtraction of a continuum model clearly reveals the  H$_2$, PAH, 
and other emission lines (Per A, Cyg A, and Cen A Figs. 5-7).

\section{H$_2$ Luminosities, Temperatures, Masses, and Surface Densities}

In order to compute the mass of warm molecular gas, it is first necessary to know the H$_2$ excitation mechanism.  The low critical 
densities (H$_2$-H$_2$ $n_\mathrm{crit} < 10^3$ cm $^{-3}$ at $T>100$ K) of the H$_2$ 0-0 S(0) and S(1) emission lines \citep{ldf99, wgf07} 
ensure that their upper levels will be collisionally excited in almost any environment where warm H$_2$ is found. Nonthermal excitation by UV 
photons or cosmic-ray secondary electrons \citep{ffh08} may contribute to the higher rotational and ro-vibrational levels, but are inefficient 
at exciting the lower pure-rotational levels. Therefore the lowest H$_2$ energy levels, which dominate the warm H$_2$ mass budget, must be in 
thermal equilibrium.

Assuming thermal equilibrium in gas at a single temperature $T$, the H$_2$ luminosity is determined by the H$_2$ mass 
$M_{\mathrm{H}2}$ and $T$ according to

$L(\mathrm{H}_2) = {M_{\mathrm{H}2} \over 2 m_\mathrm{p}} l_{\mathrm{H}2}(T)$

\noindent where the mean luminosity of a single H$_2$ molecule, summed over all transitions $u \rightarrow l$ is

$l(\mathrm{H}_2)  = \sum_{u,l} h\nu_{ul} A_{ul} f_{u}(T)$

\noindent The occupation number $f_{u}(T)$ of each level is determined by the Boltzmann 
factor and partition function. For example, a molecular gas mass of $M_9 = M_{\mathrm{H}2}/10^9 M_\odot$ and a temperature of 
200 K gives 

$L(\mathrm{H}_2) = 1.1 \times 10^{42}$ erg s$^{-1}$ $M_9$

\noindent For multiple temperature components, the luminosity of each component can be computed separately.

For all radio galaxies with two or more H$_2$ emission lines detected, we construct excitation diagrams to determine the temperature distribution 
and mass of H$_2$.  For the limiting case of an H$_2$ source size of 3\farcs7 $\times$ 3\farcs7 uniformly filling 
the width of the SL slit, we plot the column density of the upper level divided by the statistical weight for each transition 
(Figs. 8-10).  

We simultaneously fit the data with 3 temperature components (Table 11), the minimum number necessary to fit the emission line fluxes and 
parameterize what is likely a continuous temperature distribution.  We quote $1\sigma$ single parameter uncertainties for the H$_2$ temperatures 
and masses. Upper limits are included in the fits as 2$\sigma$ detections with 1$\sigma$ errorbars. This allows us to give conservative upper limits 
for the lowest temperature component for sources where the S(0) line is not detected, without affecting much the other components. Note that the 
lowest temperature component, which is constrained by the S(0) and S(1) lines dominates the total H$_2$ mass. We give an upper limit for the warmest
temperature component mass in 3C 338, where the S(6) and S(7) lines are not detected.

We assume that the H$_2$ level populations and ratios of ortho/para H$_2$ are in local thermodynamic equilibrium (LTE) and adjust the statistical 
weights of the ortho transitions iteratively to obtain a self-consistent fit. There is no evidence for non-equilibrium ortho/para ratios, which 
would manifest as suppressed or enhanced emission from odd-$J$ relative to even-$J$ transitions. The ortho/para ratios may depend upon the detailed 
thermal history of the gas, and can take much longer than the gas cooling time ($\sim 10^4$ yr) to reach their equilibrium values \citep{wcp00}. 
However, for the warm ($T>100$ K) H$_2$ observed by {\it Spitzer}, the equilibrium ortho/para ratio takes on a narrow range of values (1.6-3.0) 
such that the equilibrium assumption has only minor consequences for our model fits. For example, assuming an ortho/para ratio of 3.0 rather than 
2.1 decreases the temperature of the coolest H$_2$ component in 3C 326 from 110 to 100 K, and raises the mass by only $27\%$  to 
$2.8 \pm 0.6 \times 10^9 M_\odot$.

To determine the sensitivity of our model fits to the number of temperature components, we tried both 2-component and 3-component fits to all of the 
lines in the 3C 326 N excitation diagram. The 2-component fit gives a mass of 
$8.7 \pm 3.2 \times 10^8 M_\odot$ at 140 K ($\chi^2/\mathrm{DOF}=36.6$), compared to 
$2.2 \pm 1.2 \times 10^9 M_\odot$ at 110 K ($\chi^2/\mathrm{DOF}=1.8$) for the 3-component fit. The 3-component fit has much smaller residuals and is 
clearly better. The additonal temperature component increases the estimated H$_2$ mass by a factor of 2.5 in this case. Which lines are detected and 
used in the fit can also change the estimated H$_2$ mass. If we throw out the S(5), S(6), and S(7) lines and refit the 2 component model, we find  
$1.6 \pm 0.7 \times 10^9 M_\odot$ at 120 K ($\chi^2/\mathrm{DOF}=1.9$).

We measure radio galaxy warm H$_2$ masses from $8\times 10^6 M_\odot$ up to an amazing $2\times 10^{10} M_\odot$, the bulk at temperatures 
of 100-200 K (Tables 11 and 12). Such large H$_2$ masses imply the wholesale heating of an entire galaxy's worth of molecular 
gas in some cases. Smaller masses ($3\times 10^4-8\times 10^{7} M_\odot$) of warmer H$_2$ at 200-1500 K are also found in all galaxies where
H$_2$ is detected. Comparable luminosity is emitted by each temperature component in a given galaxy, though most of the mass resides in the 
coolest component (Table 11). In all cases where there are several high-level lines detected, the H$_2$ temperature increases with upper-level 
excitation temperature. This produces the characteristic curves in Figures 8-10, which have less negative slopes at higher energies. This 
behavior is also seen in the excitation diagram for the Perseus A filaments \citep{jhf07}. This is expected for any astrophysical situation 
where there is a range of temperatures and densities present \citep{ffh08}. 

The total warm H$_2$ column densities and column density lower limits range from $<4\times 10^{20}-1\times 10^{23}$ cm$^{-2}$
($3-800 M_\odot$ pc$^{-2}$). Only lower limits can be inferred for spatially unresolved H$_2$ emission. This spans the range 
observed for normal star-forming galaxies at the low end to ULIRGs at the high end \citep{gpc07}. 

CO emission has been detected by past observations \citep{ess99,llc03,sc03,ems05,oki05,sce06,sm07,ivb90,nbs10,olc10} of several radio MOHEGs 
(Table 12). The mass of cold ($10-50$ K) H$_2$ estimated assuming the standard Galactic CO to H$_2$ conversion factor of 
4.6 $M_\odot/$(K km s$^{-1}$ pc$^2$) ranges from $4\times 10^6-2\times 10^{10} M_\odot$. However, we caution that the applicability of the Galactic 
conversion factor to the extreme conditions in radio MOHEGs has not been established. This caveat notwithstanding, the inferred ratio of warm 
($T>100$ K) to cold H$_2$ in these galaxies ranges from $<0.18$-2.1. The galaxies 3C 218, 272.1, 326, 424, and 433 are particularly remarkable in 
having warm/cold H$_2$ ratios of $>0.5$. A large fraction of the H$_2$ in these galaxies is disturbed and heated to temperatures $>100$ K.

\section{24 $\mu$m Dust Emission}

The formation and shielding of H$_2$ requires the presence of dust, which may reveal itself through
IR continuum and PAH emission features \citep{gb10}. However, there must be a sufficient UV radiation field to heat
the very small dust grains (VSGs) and excite the PAH stretching or bending modes in order to observe 
mid-IR dust signatures. A UV photon field may be sustained by 1) star formation, 2) AGN emission, or 3) 
thermal bremsstrahlung and line emission from radiative shocks \citep{ds96}. It must be kept in mind that the H$_2$, VSG, 
and PAH emission regions are spatially unresolved and may or may not be coincident. Also, different types of sources can 
contribute in different ratios to each of these spectral features.

First we consider the ratio of H$_2$ to 24 $\mu$m dust emission (Fig. 11) for star-forming galaxies \citep{rhh07}. 
There is a trend (with large scatter) for $L$(H$_2$)/$L_\mathrm{24}$ to decrease with $L_\mathrm{24}$.  
ULIRGs \citep{hah06} have smaller $L$(H$_2$)/$L_{24}$ than nearby spirals and dwarfs, by a factor of $\sim 5$. 
This is similar to the known trend of decreasing  $L$(PAH)/$L_\mathrm{24}$ with $L_\mathrm{24}$ observed for starbursts 
and ULIRGs \citep[][and Fig. 12]{das07,tlg01}. It is debated whether this latter trend is intrinsic to star-forming regions 
or whether it reflects an increasing AGN contribution to the continuum at higher luminosity. In either case, the tendency of 
H$_2$ and PAH emission to decrease together relative to $L_{24}$ may suggest that they are both powered by star formation 
in starbursts and ULIRGs. On the other hand, \cite{z10} finds that H$_2$ emission is less obscured than PAH emission in ULIRGs,
indicating a different spatial distribution.

The radio galaxies in our sample have up to 20 times larger ratio of $L$(H$_2$)/$L_{24}$ than do normal
star-forming galaxies (Fig. 11).  In some cases (e.g., 3C 326 and Stephan's Quintet), $>10\%$ of the bolometric IR
luminosity comes out in the H$_2$ lines \citep{oaa07}. This dramatically indicates that the H$_2$ emission is not powered by star 
formation, and that H$_2$ emission is an important and perhaps primary coolant for the ISM in these sources. Strong
H$_2$ emission is often an indicator of magnetic (C-type) shocks in Galactic sources such as Herbig-Haro jets \citep{nms06} 
and supernovae \citep{ccp99,nhk07}. A deficit of 24 $\mu$m continuum relative to H$_2$ emission indicates the importance 
of non-radiative (e.g., shock) heating.

The SINGS AGNs also have enhanced $L$(H$_2$)/$L_{24}$ \citep{rhh07}, though they are not as well separated
from the star-forming galaxy locus as radio galaxies (Fig. 11). Non-radiative heating of H$_2$ appears to be important in these
nearby, radio-quiet AGNs. A similar enhancement in  $L$(H$_2$)/$L_{24}$ is seen for some of  the \cite{kos08} 'dusty ellipticals',
which are RSA ellipticals detected by IRAS \citep{gj95}. Interestingly, 12/14 of the H$_2$ detected ellipticals in the 
\cite{kos08} sample have radio sources detected in the NVSS or SUMSS survey--7 compact sources and 5 FR Is.
Radio jet activity could be responsible for exciting the excess H$_2$ emission in these sources.

\section{PAH Emission}

PAH emission features are detected in all of the radio MOHEGs except Cyg A (Table 5) and in half of the H$_2$ nondetected radio
galaxies (Table 6). However, the ratio of 7.7 $\mu$m PAH to 24 $\mu$m continuum emission  $L(\mathrm{PAH}7.7)/L_{24}$ is generally 
lower by a factor of 10-100 compared to normal star-forming galaxies from the SINGS sample (Fig. 12). The mid-IR continuum of most 
radio galaxies and radio MOHEGs is therefore not star-formation dominated. In many of these galaxies, the continuum has been shown 
to be AGN dominated, including contributions from AGN-heated dust and synchrotron emission from the jet \citep{owa06, clm07, tdh07, hwh08, loa09}. 

The large scatter that we see in $L(\mathrm{PAH}7.7)/L_{24}$ (2 dex) primarily reflects varying star formation vs. AGN fractional 
contributions to the spectra. Additional scatter is introduced by variation in the 7.7/11.3 $\mu$m PAH band ratio (Fig. 13). 
Low  $L(\mathrm{PAH}7.7)/L_{24}$ is also observed in the \cite{kos08} dusty ellipticals, indicating that the warm dust continuum in these 
sources does not come from star formation.

The median PAH(6.2 $\mu$m)/PAH(7.7 $\mu$m) ratio is 0.25 for radio MOHEGs, which is indistinguishable from the median ratio (0.28) for 
normal star-forming SINGS galaxies (Fig. 13). The median PAH(7.7 $\mu$m)/PAH(11.3 $\mu$m) ratio of 1.8 is significantly lower in radio MOHEGs 
than the median value of 4.1 measured in star-forming SINGS galaxies and within the range of values reported for SINGS AGNs \citep{sdd07}. 
The \cite{kos08} dusty elliptical galaxies have a similar median  7.7 to 11.3 $\mu$m band ratio of 1.4. \cite{sdd07} present an extensive 
discussion of possible interpretations of the observed differences in 7.7 to 11.3 $\mu$m band ratio between AGNs and star-forming 
galaxies and among AGNs.  

The impact of the PAH ionization state and size distribution on PAH emission spectra has been quantified in models 
\citep{btb01,dl01,fbv06}. In these models, the  7.7 to 11.3 $\mu$m band ratio is mainly sensitive to the ionization state and is 
lower for neutral PAHs than for cations. This interpretation is supported by observations of photodissociation regions (PDRs) which show that 
the  7.7 to $11.3\,\mu$m band ratio decreases with increasing depth into the cloud \citep{rjb05}. The ionization states of PAHs depend on the 
PAH ionization parameter $G_{UV}\times T_e^{0.5}/n_e$, where $G_{UV}$ is the $912<\lambda<2400$~\AA~ UV field energy density 
relative to the Solar Neighborhood value \citep{hab68}, and $T_e$ and $n_e$ the temperature and electron density. The electron density 
is that of the neutral, molecular, and atomic gas where PAHs are present. 

The interpretation of Galactic data leads to equal fractions of neutral and cation PAHs in the diffuse Galactic ISM for a mean ionization 
parameter $\rm \sim 1000\, K^{0.5} \, cm^{3}$ and a 7.7 to 11.3 $\mu$m band ratio (measured with PAHFIT) of $2.9 \pm 0.2$ \citep{fbv06}. 
The median value of the 7.7 to 11.3 $\mu$m band ratio of radio MOHEGs (1.8) is smaller than the value for the Galactic diffuse 
interstellar medium. This implies a smaller PAH ionization parameter, which could be accounted for by lower $G_{UV}$ in galaxies with 
a low star formation rate, or by larger $n_e$ from increased ionization by cosmic rays or X-rays.

In the Stephan's Quintet intergalactic shock, the observed PAH emission is consistent with the mass of 
warm H$_2$ gas heated by a radiation field with $G_{UV}=1.4$, assuming a Galactic dust-to-gas mass ratio and a galactic dust size distribution 
\citep{gb10}. This shows that PAHs can survive in dense regions of the interstellar medium which are shocked at velocities $<100$ km s$^{-1}$. 
PAHs could potentially be destroyed by faster shocks, which may be present in the diffuse ISM \citep{mjt10a}. In the hot $10^6-10^7$ K X-ray 
emitting gas phase, PAHs would mostly be destroyed by electron collisions and would not survive more than a few $10^3$ years \citep{mjt10b}. 
PAHs are likely protected from electron sputtering in the denser, colder phases.

We similarly expect to see PAH emission from the large H$_2$ reservoir in radio MOHEGs, excited by the ambient UV radiation field. PAH destruction 
by X-rays may possibly reduce PAH abundances in the vicinity of the AGN \citep{v91}. However, the four radio MOHEGs with significant star formation 
(discussed in \S 10--3C 31, 293, 315, and Hydra A) and correspondingly strong UV radiation field show nearly normal 7.7 and 11.3 $\mu$m PAH equivalent 
widths and PAH band ratios, demonstrating that PAH abundances have not been depleted by the AGNs on a galactic scale. The remaining radio MOHEGs show 
relatively weak PAH emission, consistent with a weaker UV field.

\section{H$_2$ to PAH Emission Ratio}

We plot the ratio of H$_2$ luminosity in the 0-0 S(0)-S(3) lines over $L(\mathrm{PAH}7.7)$ versus $L_{24}$ 
(Fig. 14). Normal star-forming galaxies from the SINGS survey \citep{rhh07} have a median ratio 
$L$(H$_2$)/$L(\mathrm{PAH7.7})=0.014$, which appears to be independent of $L_{24}$ over four decades 
($L_{24}=10^6-10^{10} L_\odot$). The mean $L$(H$_2$)/$L(\mathrm{PAH7.7})$ has a similar but slightly greater value for ULIRGs 
\citep{hah06,das07}. 

In contrast, all of the H$_2$ detected radio galaxies have very large $L$(H$_2$)/$L(\mathrm{PAH7.7})=0.03-4$ or greater. The most extreme sources, 
with $L$(H$_2$)/$L_\mathrm{PAH} > 3$ are 3C 326 N \citep{oaa07} and 3C 424. Other galaxies of different types, including the LIRG NGC 6240 
\citep{lsg03}, the Zw 3146 BCG \citep{e06}, and the Stephan's Quintet intergalactic shock \citep{a06} are also unusual in having large 
$L$(H$_2$)/$L_\mathrm{PAH}$ ratios. Strikingly, all of the SINGS Seyferts and all but 4 of the SINGs LINERs \citep{rhh07} also stand out as 
having relatively large $L$(H$_2$)/$L_\mathrm{PAH}=0.02-0.2$, though not so large as the most extreme radio galaxies. Such large ratios 
indicate shock, cosmic ray, or X-ray heating rather than UV heating of the molecular gas phase \citep{rkl02,rhh07}. We explore the various 
possible heating mechanisms for H$_2$ in radio galaxies below. 

Alternatively, it might be argued that a high ratio $L$(H$_2$)/$L(\mathrm{PAH7.7})$ is caused by a low abundance of PAHs in the warm molecular gas 
rather than low UV field. A UV field of $G_{UV} \sim 4$ would be required explain the 11.3 $\mu$m PAH flux from 3C 326 N if it is associated with the 
observed warm H$_2$ and CO \citep{nbs10}, assuming a normal H$_2$/CO conversion factor (Table 12). We use archival GALEX UV data to independently
constrain the intensity of the radiation field and PAH abundance. The NUV AB magnitude is 21.47, which translates into a surface brightness of 
$1.59\times 10^{-6}$ W m$^{-2}$ sr$^{-1}$, assuming emission from star-forming regions uniformly covering a 5 arcsec (8 kpc) diameter area. This 
corresponds to a mean UV radiation field $G_{UV}$ = 12. If any of the observed UV radiation comes from the AGN, the $G_{UV}$ value contributing to 
PAH excitation would be less.  A comparison of the two $G_{UV}$ estimates gives an upper limit on the PAH depletion of $\sim 3$. This is far from 
sufficient to explain the extreme ratio $L$(H$_2$)/$L(\mathrm{PAH7.7})>4$ in 3C 326 N, which is 300 times the mean value in normal star-forming 
galaxies.

\section{MOHEG Definition}

Since we do not have far-IR measurements for most of the sources in our sample, we do not use our original MOHEG definition \citep{oaa07}.
We could re-express that definition as $L$(H$_2$)/$L_{24}>2\times10^{-2}$ (dashed line in Fig. 11), excluding all normal star-forming galaxies 
from the MOHEG class. The observed H$_2$ to continuum contrast ratio for our H$_2$ detected radio galaxy sample ranges from 
$L$(H$_2$)/$L_{24} = 6\times 10^{-4}-0.4$, 
which means that only 9/17 of these galaxies would be considered MOHEGs by such a definition.  Inspection of Figure 11 suggests that defining 
MOHEGs in terms of a minimum $L$(H$_2$)/$L_{24}$ contrast ratio is problematic. For one, this ratio decreases with $L_{24}$ for normal star-forming
galaxies, so a line with negative slope in that figure would do a better job at separating MOHEGs. However, it would not completely account for 
the variation in AGN continuum contribution at 24 $\mu$m, which introduces scatter into this relation. 

The much cleaner separation between MOHEGs and normal star-forming galaxies in Figure 14 compared to Figure 11 motivates us to 
redefine MOHEGs as galaxies with $L$(H$_2$)/$L(\mathrm{PAH}7.7)> 0.04$. This particular cutoff was chosen to exclude all non-AGN SINGs galaxies from 
the MOHEG class. With our new definition, 16/17 of the H$_2$ detected radio galaxies, 10/18 of the  \cite{kos08} dusty ellipticals, NGC 6240, and 
the Zw 3146 BCG are all MOHEGs.  We call radio galaxies that are MOHEGs 'radio MOHEGs' to distinguish them from MOHEG samples selected in other 
wavebands. The radio galaxy 3C 31, with  $L$(H$_2$)/$L(\mathrm{PAH}7.7)=0.03$ falls just below the MOHEG cutoff as a result of the relatively large 
contribution of star formation and strong 7.7 $\mu$m PAH emission its mid-IR spectrum (Fig. 1). However, for simplicity and since it has a larger 
$L$(H$_2$)/$L(\mathrm{PAH}7.7)$ ratio than any SINGs galaxy of similar luminosity $L_{24}$, we will also call this source a radio MOHEG. 
It is not possible with the current data to reliably classify the H$_2$ nondetected radio galaxies with respect to the MOHEG criterion. 
Therefore, the 31\% MOHEG fraction in our radio galaxy sample should be considered a lower limit.

The success of the  $L$(H$_2$)/$L(\mathrm{PAH}7.7)$ diagnostic ratio in distinguishing between
AGN and star-formation dominated spectra reflects the different molecular gas heating mechanisms in the two classes. Furthermore,
the $L$(H$_2$)/$L(\mathrm{PAH}7.7)$ value at a given $L_{24}$ might be used to determine the ratio of AGN to star-forming 
contributions to these two emission features. If the H$_2$ and 7.7 $\mu$m PAH emission have different spatial extent, this ratio and classification 
as a MOHEG might depend upon the aperture used to extract the spectrum. There is a similar issue with AGN optical classification,
since AGN fraction tends to increase with decreasing aperture size.

\section{Star Formation}

In general, 3C radio galaxies tend to have less luminous PAH emission and lower inferred star-formation rates than other AGN samples, 
including PG quasars and 2MASS red quasars \citep{sor07}. Whether this is separately true for those with and without hidden quasars 
\citep{owa06} needs to be investigated. For only 4/17 radio MOHEGs do the spectra (Fig. 1) indicate a large star formation fraction 
(3C 31, 218, 293, and 315), with  $L(\mathrm{PAH}7.7)/L_{24}>0.1$ (Fig. 12). Based on template fitting, the star formation contribution in 3C 31, 
218 and 293 ranges from  87-100\% at 15 $\mu$m \citep{loa09}. Using the same procedure, we find that 3C 315 has a 40-50\% star formation 
contribution at 15 $\mu$m.

We use the observed 7.7 $\mu$m PAH luminosity to estimate star formation rates (SFRs, Table 13) for radio MOHEGs using the 
prescription of \cite{rsv01}. We convert the 11.3 $\mu$m PAH luminosity to 7.7 $\mu$m PAH luminosity, dividing by 
the mean 11.3/7.7 $\mu$m PAH ratio of 0.26 \citep{sdd07} to make an equivalent SFR estimate for sources where the 7.7 $\mu$m PAH is not 
detected. Note that the SFRs are integrated over the {\it Spitzer} IRS slit, which does not cover the whole galaxy for most of the 
observations (Figs. 1-4). The SFRs for MOHEGs with detected  7.7 $\mu$m PAH emission range from $4\times 10^{-3} M_\odot$ yr$^{-1}$ for 3C 272.1 at 
the low end to $2.6 M_\odot$ yr$^{-1}$ for 3C 436 on the high end. Since the 11.3/7.7 $\mu$m PAH ratio is systematically enhanced compared to
normal star-forming galaxies (Fig. 13), the 11.3 $\mu$m PAH star-formation rates should be considered upper limits and used with caution.

\section{AGN Activity}

Most (13/17) of the radio MOHEGs are mid-IR weak \footnote{The wavelength used in the \cite{owa06} definition of MIR-weak 
($\nu L_\nu(15 \mu\mathrm{m}) <8 \times 10^{43}$ erg s${-1}$ $=2\times10^{10} L_\odot$) is different than the one we employ here, 
because of the need to accomodate a different redshift range.} \citep{owa06,loa09}, with $\nu L_\nu(24 \mu\mathrm{m}) <2\times 10^{10} L_\odot$ 
(Table 9). Most are also optically classified as low-ionization galaxies (LIGs, Table 1) based on their low [O{\sc iii}] 5007/[O{\sc ii}] 3727 ratios. 
Using the optical emission line measurements of \cite{bcc09}, we construct BPT diagnostic diagrams \citep{bpt81}, which show that most radio MOHEGs 
are LINERs according to the \cite{kgk06} line ratio criteria (Fig. 15). We can not say whether this tendency of radio MOHEGs to host MIR-weak, LINER 
AGNs is intrinsic or rather represents a bias against detecting low-equivalent width H$_2$ emission lines and PAH features against the stronger 
continuum in the MIR-luminous HIGs and BLRGs.

The mid-IR forbidden line ratios of radio MOHEG LINERs (Fig. 16) are similar to those of other LINERs observed by {\it Spitzer}, 
with  [Ne {\sc iii}]/[Ne {\sc ii}]$<2.0$, [O {\sc iv}]/[Ne {\sc ii}] $<1.0$, [S {\sc iii}]/[Ne {\sc ii}]$<1.0$, and
[S {\sc iii}]/[Si {\sc ii}]$<0.5$  \citep{src06,dsa06,dsm09}. The low [S {\sc iii}]/[Si {\sc ii}] ratio in particular is
characteristic of AGN X-ray dissociation regions \citep{dsa06}, in contrast to the higher ratio typical of star-forming 
galaxies. The [Ne {\sc iii}]/[Ne {\sc ii}] and [S {\sc iii}]/[Si {\sc ii}]  ratios of radio MOHEGs are however considerably greater than what are 
seen in the Stephan's Quintet main shock \citep{ca10}. This may indicate that radio MOHEG forbidden line emission is associated primarily with the 
AGN rather than the shocked H$_2$ emission region. On the other hand, we can not rule out the possibility of a significant contribution from fast 
J-type shocks (with similar line ratios to LINERs) to the optical and MIR forbidden emission lines, \citep[e.g., 3C 326,][]{nbs10}.  

The three radio MOHEGs with Seyfert-like, high-ionization optical spectra (Cyg A, 3C 433, and 436) all have 
[Ne {\sc iii}]/[Ne {\sc ii}]$>2.0$ and [O {\sc iv}]/[Ne {\sc ii}] $>2.0$. High ionization [Ne {\sc v}] and [Ne {\sc vi}] lines  
are detected in Cyg A and 3C 433, but only the [Ne {\sc vi}] 7.65 $\mu$m line is detected in 3C 436 (Table 8). 
All three host powerful, hidden quasars with $\nu L_\nu(24 \mu\mathrm{m}) > 2 \times 10^{10} L_\odot$, and deep silicate troughs 
($\tau_{9.7}>0.5$) indicate obscuration of the AGN by additional cold dust \citep{owa06}.  Polarized, scattered broad emission lines are seen in the 
optical spectra of Cyg A \citep{ocm97} and 3C 433 (Ogle et al., in preparation), confirming the presence of hidden quasar nuclei. The [Ne {\sc v}] and 
[Ne {\sc vi}] lines are detected at 2-3$\sigma$ in Cen A \citep[see also][]{whh05,mls99}, which may contain a hidden Seyfert nucleus 
\citep{wa04,loa09}.

\section{Insufficiency of AGN X-ray Heating as an H$_2$ Power Source}

The lack of a high-luminosity, high-ionization AGN in most radio MOHEGs indicates weak accretion onto the central supermassive black hole.
Such low luminosity AGNs are insufficient to power the H$_2$ emission via X-rays. 

The process of H$_2$ heating by X-rays in X-ray 
dissociation regions (XDRs) is described by \cite{mht96}. In their XDR models, 30-40\% of the absorbed X-ray flux goes into gas heating 
via photoelectrons.  The atomic photoelectric cross section energy dependence is $E^{-8/3}$, so that photoionizations in the energy range 
1-30 keV are important. For a standard AGN X-ray power law spectrum with frequency dependence $\nu^{-0.7}$, 
$L({\rm 1-30 keV})/L({\rm 2-10 keV)}=2.3$. 

For a characteristic observed H$_2$ temperature of 200 K (corresponding to an
ionization parameter $\log \xi_\mathrm{eff} = -2.6$) the cooling by H$_2$ rotational lines in XDR models is $\sim 2\%$ of the total gas 
cooling \citep[][Figs. 3a,5]{mht96}. At this temperature, the ratio of the first four rotational lines to the total rotational line luminosity 
is  $L$(H$_2$ 0-0 S(0)-S(3))/$L$(H$_2$)$=0.58$. Combining the above factors, we estimate a maximum H$_2$ to X-ray luminosity ratio of 
$L$(H$_2$ 0-0 S(0)-S(3))/$L_\mathrm{X}$(2-10 keV)$<0.01$. This ratio is conservative, since it assumes that all of the X-ray flux 
from the AGN is absorbed by the XDR.

We compare unabsorbed nuclear X-ray (2-10 keV) luminosities measured by {\it Chandra} (Table 9) with the summed H$_2$ S(0)-S(3) 
line luminosities measured by {\it Spitzer} (Fig. 17). Two sources lack X-ray nuclear flux measurements (3C 310 and 424). 
We find $L$(H$_2$)/$L_\mathrm{X}$(2-10 keV)$>0.01$ in all measured radio MOHEGs and SINGS AGNs, except Cyg A. The warm H$_2$ in Cyg A would have 
to intercept and absorb $>30\%$ of the X-ray luminosity from the AGN to produce the observed mid-IR H$_2$ line luminosity.  For all other sources, 
the observed  rotational $L$(H$_2$)/$L_\mathrm{X}$(2-10 keV) exceeds the maximum XDR H$_2$ to X-ray luminosity ratio estimated above. The large 
scatter (nearly 5 dex) in the observed ratio  $L$(H$_2$)/$L_\mathrm{X}$(2-10 keV)$=0.002-100$ also argues strongly against X-ray heating as the 
primary driver of H$_2$ emission in radio galaxies or SINGS AGNs.

\section{Host Galaxies and Environs}

Many radio MOHEGs (Table 3 and Figs. 1-4) appear to have a galactic stellar disk \citep{dcm07}, dust disk, or dust lanes \citep{dbb00}. 
Stellar or dust disks may indicate a past gas-rich merger, which would help to explain the presence of large quantities of H$_2$.  Most of 
the other galaxies are peculiar or distorted in their stellar isophotes. Only 3C 386 appears to have a regular elliptical host 
morphology.

There is evidence for patchy or organized dust absorption in 12/17 radio MOHEGs (Table 3).
This is important because H$_2$ forms most easily in the presence of dust. Dust masses measured from HST optical images
\citep{dbb00} are typically $<1\%$ of the warm H$_2$ mass measured with {\it Spitzer} (Table 12). However, this method 
likely underestimates the total dust mass associated with the warm H$_2$ and cold CO. The hosts of 3C 31 and 270 
have striking dust disks and Cen A has thick dust lanes from an edge-on warped disk. Dust lanes, patchy dust, or 
dust filaments are also seen in Per A, 3C 272.1, 293, 315, 317, 338, Cyg A, 3C 433, and 436. The only galaxies without 
obvious dust absorption are Hydra A, 3C 310, 326 N, 386, and 424. However, dust absorption could be missed in the HST 
snapshots because of a smooth dust distribution or low galaxy inclination.

Interestingly, 14/17 radio MOHEGs belong to close galaxy pairs, groups or clusters (Table 3). Of
these, 3 reside in pairs, 3 in groups, 1 in a poor X-ray detected cluster, and 7 in X-ray luminous clusters. Four of the 
clusters (hosting Per A, Hydra A, 3C 317, and 3C 338) are so-called cool-core clusters with $\sim 3$ keV gas temperatures and high 
central X-ray surface brightness. The cooling flow phenomenon may be a way to deliver large quantities of gas to these radio MOHEGs.
FR I radio galaxies tend to reside in rich environments at low redshift \citep{ls79, hcb86}, while low-$z$ FR II radio galaxies prefer 
the field. Because radio MOHEGs are found in both environments, cluster membership is apparently not a prerequisite for the H$_2$ 
emission phenomenon.  Gravitational interactions with nearby companions may play a significant role in delivering or driving the molecular 
gas into the center of radio MOHEGs in the field or in clusters. A statistical comparison of the frequency of physical galaxy pairs in the 
H$_2$ detected and nondetected radio galaxy subsamples is outside the scope of this paper.

There are tidal distortions indicating strong interactions in at least 4/7 galaxy pairs, groups, and poor clusters:
3C 293, 310, 326, and 433. The 3C 31 and 3C 315 pairs do not show strong distortion, but the halos of close
companions appear to overlap the primary galaxy. The Virgo cluster galaxy 3C 270 (NGC 4261) has a faint tidal tail \citep{tvn09}.
The 4 cool-core cluster cDs and 3C 433 have large, diffuse halos which appear to envelop two or more galaxies in the cluster centers. 
The three galaxies that do not reside in pairs, groups or clusters are 3C 386, 436, and Cen A. The warped, dusty molecular disk in Cen A 
may be the result of an elliptical/spiral merger \citep{bm54,qbk06}. The host of 3C 436 shows signs of a recent merger, including a second 
nucleus and other irregularities in its surface brightness \citep{mkd99}.  The Virgo cluster galaxy 3C 272.1 (M84), the group galaxy 3C 424, 
and the isolated galaxy 3C 386 show no obvious signs of recent interaction.

\section{Radio Source Morphologies}

The radio MOHEGs in our sample cover a large range of radio luminosity  $\nu L_\nu(178 \mathrm{MHz})=10^{39}-10^{44}$ erg s$^{-1}$ 
(Table 9 and Fig. 18). They display a motley collection of radio morphologies (Tables 1 and 2). There are 6 FR IIs and 11 FR Is. The fraction of 
FR II MOHEGs is comparable to the fraction of FR I MOHEGs in the sample ($22 \pm 9 \%$ vs. 39 $\pm$ 12 \%). However, there could 
potentially be a bias against detecting H$_2$ in FR IIs with hidden quasars because the emission lines might be overwhelmed by copious 
warm dust emission.  

Further classification of radio morphologies is somewhat subjective but may offer clues to the origin of radio MOHEGs.
Looking at their morphology subclassifications (Table 1), 5/11 FR Is are classical twin jet (TJ) sources, 4/11 are fat doubles 
(FD), 1/11 is a double-double (DD), and  1/11 has a compact symmetric core (CSC). Of the FR IIs, only 3/6 are classical doubles, 
1/6 is a fat double (FD), and 2/6 are X-shaped (X). There is a relatively large fraction of distorted radio sources compared to the 
rest of the sample, which only has 2/38 FD and 0/38 X-shaped sources. The two X-shaped radio sources (3C 433 and 3C 315) and 
the very peculiar source 3C 338 may be distorted by the motions of their interacting hosts or have jets deflected by their companions.

The radio MOHEGs also include a high fraction (3/17) of so-called restarting double-double (DD) or CSC sources, 
compared to 1/38 of the H$_2$ nondetected radio galaxies. The radio map of 3C 424 shows a brighter inner set of lobes 
surrounded by a fainter set of lobes that appears to originate from a previous spurt of activity \citep{bbl92}. High resolution 
maps of 3C 293 and 315 reveal bright, inner, compact symmetric double radio lobes roughly aligned with the larger scale 
radio lobes.  These may represent new outbursts of radio activity or interaction of the small scale radio source with the host galaxy 
ISM. The radio cores also appear relatively bright in 3C 310, and 386, perhaps also indicating rejuvenated jets, or else
high Doppler beaming factor. The restarting jet phenomenon may be connected to a new supply of fuel provided by the large mass of 
H$_2$ in these galaxies. Interaction between the restarting jet and the host galaxy ISM may play an important role in heating the H$_2$.

\section{Discussion}

\subsection{Radio-Jet Feedback}

The high incidence of MOHEGs in our radio-selected sample may indicate that the radio jets or lobes power the H$_2$ emission.
However, there is at best a weak correlation between radio MOHEG H$_2$ luminosity and 178 MHz radio lobe power $L_{178}$ (Fig. 18a). The linear 
Pearson correlation coefficient is $R=0.79$, which reduces to $R=0.35$ ($<17\%$ probability of a chance correlation) when removing the 
partial correlation of luminosity with distance. For MOHEGs, there is a large range $L$(H$_2$)/$L_{178}=0.004-6$ in the ratio of H$_2$ luminosity to 
low-frequency radio lobe synchrotron luminosity. The upper limits for H$_2$ nondetected radio galaxies have similar scatter, with a lower median
ratio than the radio MOHEGs in this plot. The available mass of H$_2$ must be a limiting factor, and is likely an important source of scatter in 
the H$_2$ luminosity. 

Scatter in the 178 MHz radio luminosity may be caused by environmental density 
variations or spectral aging of the radio lobes.  Size and time scale differences between the radio and H$_2$ emission regions may also 
be important. For most sources, the spatially unresolved (kpc scale) H$_2$ emission region is much smaller than the radio source, which extends 
far beyond the host galaxy (50 kpc-1 Mpc). The extended radio emission may not give a reliable indicator of the jet power dissipated
at kpc scale. However, while the H$_2$ cooling time scale is only $\sim 10^4$ yr, the dissipation timescale of the molecular gas kinetic energy 
may be even greater than the typical radio source lifetime of $\sim 10^7$ yr \citep{nbs10}. Therefore, variations in jet power over the lifetime of the
radio jet may not cause significant variations in the H$_2$ luminosity.

There is no significant correlation between $L$(H$_2$) and the 5 GHz radio core power $L_5$ measured at arsecond scales (Fig. 18b). This 
is not surprising since relativistic beaming can boost or de-boost the core flux by a large factor. Differences in radio spectral shape may 
also be large at this high frequency. Neither is there a correlation of $L$(H$_2$)/$L(\mathrm{PAH}7.7)$ with $L_{178}$ nor $L_5$ (Fig. 18c,d). 
This ratio and the fraction of warm/cold H$_2$ may depend on a number of factors, including the spatial distribution of H$_2$ relative to the 
radio source, the pressure of the hot ISM, and the detailed jet-ISM interaction history.

It is instructive to compare radio MOHEGs to dusty elliptical MOHEGs selected by IRAS \citep{kos08}. We estimate 178 MHz luminosities
for 12 ellipticals by interpolating or extrapolating $100-10^3$ MHz radio fluxes measured by the NVSS, SUMSS, or FIRST surveys, or fluxes 
available in NED (Fig. 18a, c). These galaxies are up to 100 times fainter than radio MOHEGs at 178 MHz, and extend the MOHEG $L$(H$_2$) vs. 
$L_{178}$ parameter space. The faintest radio sources with $L_{178}<10^{38}$ erg s$^{-1}$ (in NGC 2974, NGC 3894, IC 3370, and NGC 5044) 
lie well above the $L$(H$_2$) vs $L_{178}$ locus established by radio MOHEGs, with $L$(H$_2$)/$L_{178}>50$. We suggest that these 
weak radio sources may be radio galaxies that are currently in between strong radio jet outbursts. This hypothesis can be tested by 
searching for low-frequency radio relics. The H$_2$ emission may be powered by residual kinetic energy injected into the ISM by previous 
radio jet activity. The H$_2$ luminosity floor of $L$(H$_2$)$\sim10^{39}$ erg s$^{-1}$ may be set by the ratio of the H$_2$ kinetic energy 
dissipation timescale and the mean time between radio outbursts. Alternatively, this low level of H$_2$ emission could be powered by gas 
accretion or X-ray emission from a weak AGN. It will be important to test the latter possibility with additional {\it Chandra} observations.

Several of the radio MOHEGs that reside in X-ray bright clusters show evidence that the radio lobes have displaced the intracluster medium
(ICM) in X-ray cavities. For these sources, it is possible to estimate the $4pV$ energy required to excavate the cavities,
and divide by the buoyancy time scale to estimate the average jet power deposited into the ICM \citep{rmn06}. These
jet cavity powers fall in the range $10^{42}-10^{45}$ erg s$^{-1}$ and the ratio of H$_2$ luminosity to jet cavity power falls in the range 
$1\times 10^{-4}-3\times 10^{-3}$ (Table 14). This demonstrates that the jet power is more than sufficient to produce the observed H$_2$ luminosity, 
modulo an unknown conversion efficiency. 

One potential difficulty is efficiently coupling the jet power to the host galaxy ISM after the radio lobes have broken out of the host 
galaxy and are doing most of their work on the intergalactic medium. \cite{oaa07} argued that only $7\times 10^{-4}$ of the radio jet power 
in 3C 326 could be dissipated in the host galaxy ISM, based on the small ratio of radio core to Mpc-scale radio lobe flux. This in turn 
could only power $<10\%$ of the observed H$_2$ luminosity, for an estimated jet kinetic luminosity of $\sim 10^{44}$ erg s$^{-1}$. Given 
the uncertainties in estimating jet kinetic luminosity and time variability, the inner radio jet could easily be ten times more powerful 
than this estimate. In addition, the radio core power is probably not a reliable tracer of jet kinetic power dissipation and may be strongly
affected by Doppler boosting or de-boosting, as indicated by the large scatter in Figure 18b. 
Turning this problem around, the large H$_2$ luminosity from the center of 3C 326 N may indicate that a significant fraction of the radio jet
power is dissipated inside the host galaxy, even though the radio lobes have long since escaped.  

On the other hand, the compact symmetric radio lobes in the radio MOHEG 3C 293 \citep{ess99, emt05, oal08, pvi10} appear to be directly 
interacting with the host ISM. It will be important to measure the pressure in the hot interstellar medium with {\it Chandra} to 
determine if dissipation of jet kinetic energy can indeed power the H$_2$ emission in this and other kpc-size sources (e.g., 3C 315).

High-velocity neutral and ionized outflows have been observed in the radio MOHEG 3C 293, and there is morphological and kinematic evidence 
that these outflows are driven by the radio jet \citep{moe03,emt05}.  Similar  H {\sc i} and ionized outflows have been found in several other 
low-redshift compact and extended radio sources \citep{mto05, htm08}, with outflow speeds of up to 2000 km s$^{-1}$ and mass-loss rates of up 
to 50 $M_\odot$ yr$^{-1}$ indicating a strong interaction between the radio jet and the host galaxy ISM. \cite{nbs10} recently discovered 
a high velocity ($\sim 1000$ km s$^{-1}$) neutral (Na {\sc I}) outflow in 3C 326, the most extreme radio MOHEG. Even more spectacular jet driven 
ionized outflows, which may be responsible for ejecting a large fraction of the host galaxy ISM, are observed in 
high-redshift radio galaxies \citep{nld08}.

The connection between jet driven outflows and H$_2$ emission in radio MOHEGs is studied in detail by \cite{nbs10}.  Since 
the molecular hydrogen must have high density and a low volume filling factor, it is likely that the cocoon of hot, shocked ISM surrounding 
the jet \citep{s74,bc89} serves as an intermediary between the jet and the H$_2$. The radio jet heats the diffuse ISM, which expands in a hot
fast wind and drives shocks into ambient or entrained molecular clouds, powering the observed warm H$_2$ emission. For H$_2$ at densities of 
$10^3$-$10^4$ cm$^{-3}$, the observed range of temperatures is characteristic of 5-20 km s$^{-1}$  shock velocities \citep{gb09}. 

The relative velocity and shear between the molecular clouds and hot wind may ultimately power the H$_2$ emission through turbulent mixing 
layers \citep{nbs10,gb09}.  The hydrodynamical stability of the H$_2$ emitting  clouds needs to be investigated, as they may be 
ablated by the hot wind and may eventually dissolve completely. However, there is evidence that the atomic and H$_2$ filaments in Per A are 
held together by strong magnetic fields \citep{fjs08}. Moreover, the fact that H$_2$ is the primary coolant only in magnetic (C-type) shocks 
strongly suggests that the magnetic field plays an important role in MOHEGs.

\subsection{Cosmic Ray Heating and Pressure}

Another potentially important source of heating for warm H$_2$ in radio MOHEGs may be ionizing particles in the form of relativistic particles 
from the radio lobes. \cite{ffh08} have proposed cosmic rays or thermal particles from the surrounding hot gas as the dominant heat source 
powering the warm H$_2$ emission filaments in the Perseus A  (3C 84) cool core cluster. An enhanced cosmic ray density has also been proposed to 
account for the high temperature of molecular clouds in the Milky Way nuclear disk \citep{ywr07}. We discuss this possibility for H$_2$ luminous 
radio galaxies. 

The critical density of the S(0) and S(1) lines is sufficiently low \cite[$<10^3$ cm$^{-3}$ at $T>100$ K]{ldf99,wgf07} that we can assume that the 
corresponding excited levels are thermalized by collisions. Thus, the line emission is cooling the gas and must be balanced by gas heating. 
In our radio MOHEG sample, the mean of total line emission per warm H$_2$ molecule (in the lowest temperature component, Table 11) is observed 
to be $\rm 7\times10^{-24}\, erg\,s^{-1}$.

The molecular gas heating  by cosmic rays is  4~eV per ionization in the MEUDON gas code \citep{lnl06}. In CLOUDY, the heating 
efficiency increases with the gas ionization fraction; the heating per ionization is 7~eV for a ionization fraction of  $10^{-4}$ 
characteristic of the diffuse interstellar medium where carbon is ionized \citep{sfa05}. The H$_2$ line cooling is 
balanced by cosmic ray heating for an ionization rate per hydrogen  $\zeta \sim 7\times 10^{-13} \, {\rm s^{-1}}$, a value $3.5\times 10^4$
times higher than the standard Galactic rate  \citep{sfa05}. For such a value, cosmic rays are the main 
destruction path of H$_2$ molecules and  the molecular gas fraction depends on the ionization rate to gas density ratio $\zeta /n_H$. 
Model calculations presented by \cite{ffh08} show that for $\zeta \sim 7\times 10^{-13} \, {\rm s^{-1}}$ the  gas is molecular for 
$n_H > 5\times 10^4$ cm$^{-3}$ (see their figure 2). Note that for such high densities H$_2$ rotational states  are thermalized up to 
$J \ge 5$, and the higher J lines up to at least S(3) are also cooling lines. 

Combining the lower limit on the gas density and the warm H$_2$ temperature ($\sim 200\,$K) inferred from the S(1) to S(0) line ratio 
we get a warm molecular gas pressure $\sim 10^7 \,{\rm K \, cm^{-3}}$ which is a factor 1400 higher than the mean cosmic ray pressure 
of $7.2\times10^3 \,{\rm K \, cm^{-3}}$ in the Milky Way disk \citep{bac90}. For a cosmic ray energy density $3.5\times 10^4$ times the Galactic 
mean value, the cosmic ray pressure would be 25 times larger than the thermal gas pressure (assuming that the cosmic ray pressure scales linearly 
with the ionization rate). Such a large difference in pressure would seem unsustainable, however it could in principle be 
supported by magnetic pressure in the compressed magnetic field within molecular clouds \citep{pbf09}.  Therefore, 
heating by cosmic rays from the radio lobes or hot gas in the radio jet cocoon may present an alternative or additional pathway to heat H$_2$ by 
radio-jet feedback.

\subsection{Gas Supply from Cooling Flows and Galaxy Collisions}

An important unanswered question is the origin of large quantities of molecular gas in radio MOHEGs. 
Copious amounts ($10^9-10^{11} M_\odot$) of cold molecular gas discovered in $\sim 45\%$ of BCGs in X-ray selected cool-core clusters 
\citep{edge01,sc03} may be deposited by cooling flows. H$_2$ emission from warm molecular gas also appears to be common in cool-core BCGs 
\citep{e06,jhf07,dm09}. Since a large fraction of these BCGs host radio sources,  the H$_2$ emission could potentially be excited by radio jet 
feedback. The gas cooling rates in the radio MOHEGs that reside in cool core clusters are $10-200 M_\odot$ yr$^{-1}$ \citep{pkp03,hm05}, enough
to supply the observed molecular gas masses in $<1$ Gyr. 

Gas-rich galaxy collisions or tidal interactions with nearby galaxies may drive gas into the center of radio galaxies in less rich environments
\citep{oaa07}. This is an attractive possibility, considering that most radio MOHEGs are mergers or have nearby companions, and several display 
prominent tidal distortions. In comparison, most field elliptical galaxies and 3C radio galaxies in the nearby universe are poor in 
molecular gas \citep[$<10^9 M_\odot$,][]{cyb07,llc03,olc10}. Galaxy collision driven inflows and jet induced outflows may constitute 
another, intermittent type of AGN feedback loop \citep[e.g.,][]{App02}.

While cooling flows or galaxy collisions may deliver the molecular gas in radio MOHEGs, it seems unlikely that gas accretion or galaxy 
collisions contribute significantly to heating the H$_2$. Unlike the galaxy-collision powered H$_2$ emission from Stephan's Quintet \citep{a06}, the 
bulk of the H$_2$ emission is concentrated inside the central $\sim 10$ kpc of radio galaxies. We are pursuing higher resolution spectroscopy
of mid-IR and near IR H$_2$ emission lines in order to determine the dynamical state of the warm molecular gas.

\section{Conclusions}

(1.) We detect high luminosity H$_2$ emission lines from 17/55 (31\% of) radio galaxies at $z<0.22$ observed with {\it Spitzer}
    IRS at low resolution. 

(2.) We redefine molecular hydrogen emission galaxies (MOHEGs) to be galaxies with $L$(H$_2$)/$L(\mathrm{PAH 7.7})>0.04$. 
     This puts 16/17 of the H$_2$-detected radio galaxies (radio MOHEGs) in this rapidly growing, new class of galaxies. 

(3.) Large H$_2$/PAH ratios indicate that the H$_2$ emission from MOHEGs is most likely powered by shocks. We can not completely rule out a 
     contribution from cosmic ray heating, though it would require cosmic ray pressures that exceed the gas pressure by a large factor.

(4.) The star formation rates in radio MOHEGs range from very low to moderate ($4\times 10^{-3} - 3 M_\odot$ yr$^{-1}$),
     and the 24 $\mu$m continuum is dominated by the AGN in 13/17 of these galaxies.

(5.) Most MOHEG AGNs in our sample are mid-IR weak and have low-ionization forbidden emission line spectra.  
     
(6.) AGN X-ray emission measured  by {\it Chandra} is not luminous enough to power the H$_2$ emission in MOHEGs.

(7.) All but one of the radio MOHEGs in our sample reside in merger remnants, interacting pairs, groups, or clusters, 
     supporting the hypothesis that galaxy interactions and mergers may deliver molecular gas or drive existing
     molecular gas into the central few kpc of radio galaxy hosts. 

(8.) Radio jet driven outflows of hot gas may drive shocks into the molecular gas, powering the H$_2$ 
     emission in radio MOHEGs. The jet power measured for radio MOHEGs in cool core clusters is more than 
     sufficient to power the H$_2$ emission.  Higher spatial resolution of the H$_2$ emission 
     region and its kinematics will give further insights into the radio jet feedback mechanism.

\acknowledgements

This work is based primarily on observations made with the {\it Spitzer} Space Telescope, which is 
operated by the Jet Propulsion Laboratory, California Institute of Technology under NASA contract 1407. 
Supporting observations were retrieved from the {\it Chandra} archive, maintained by the 
Chandra X-ray Observatory Center, which is operated by the Smithsonian Astrophysical Observatory on behalf of 
NASA. Optical {\it Hubble} images and NUV GALEX data were retrieved via MAST, operated by the Space Telescope 
Science Instute for NASA. We have made use of the NASA/IPAC Extragalactic Database (NED) which is operated by the Jet
Propulsion Laboratory, California Institute of Technology, under contract with NASA. Support for 
this research was provided by NASA through an award issued by JPL/Caltech. We thank Bill Reach for 
providing us with his H$_2$ fitting code. We also thank Simona Giacintucci for providing the low frequency
radio flux of the MOHEG elliptical NGC 5044 in advance of publication.

\clearpage

\clearpage
\begin{figure}
  \plotone{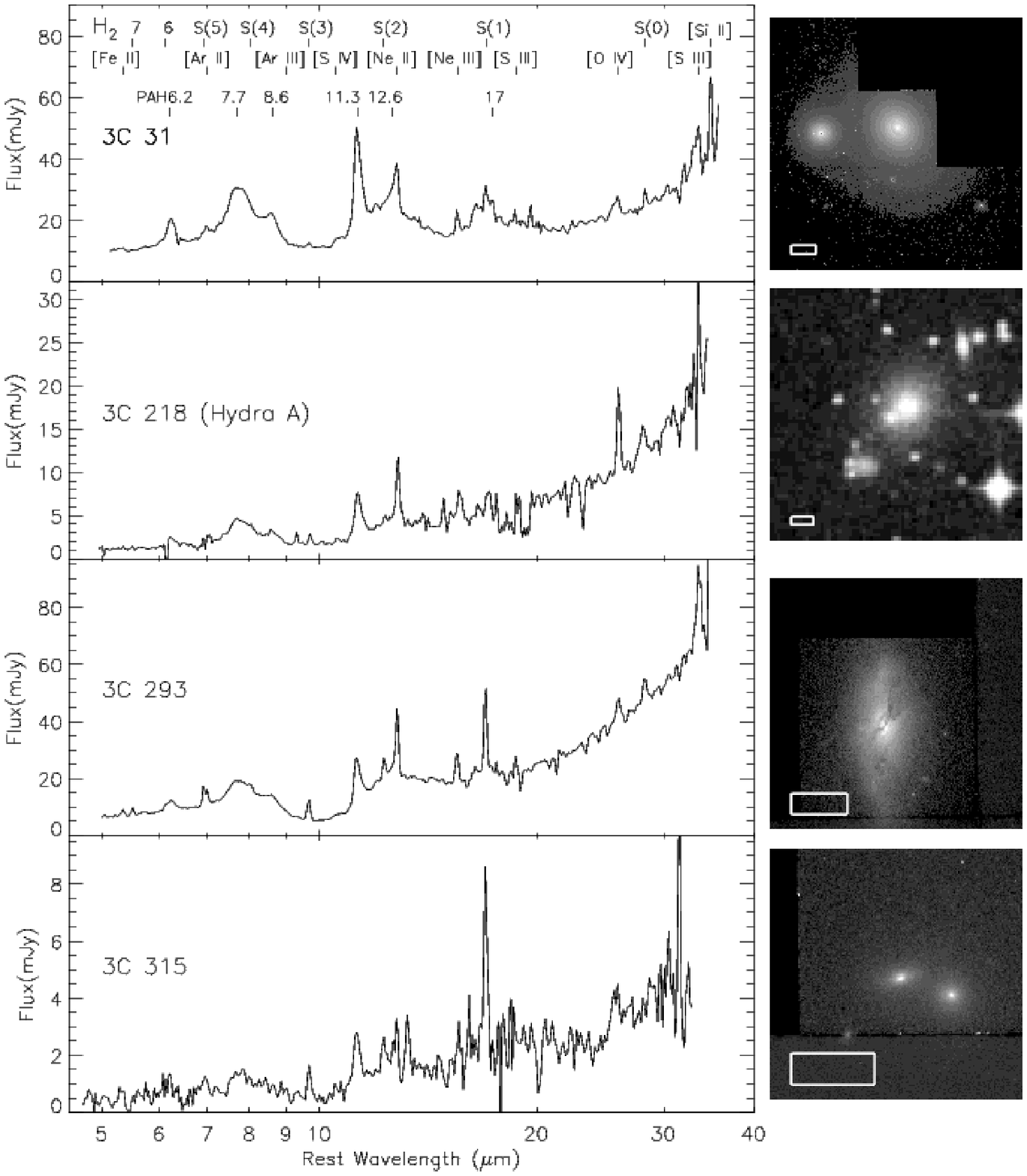}
  \caption{{\it Spitzer} IRS low-resolution spectra of 3C 31, 218 ($=$ Hydra A), 293, and 315. These sources
              have strong PAH emission and 9.7 $\mu$m silicate absorption. Spectra are arranged in order of increasing
              equivalent width of the H$_2$ pure-rotational lines. Images to the right are from HST WFPC2 (3C 31, 293 and 315) 
              and the UK Schmidt telescope (3C 218, from NED).  The host of 3C 31 resides in the Arp 331 group. Hydra A resides in the 
              cool core cluster A 780. The box at the lower left corner of each image shows the size of the 
              $3\farcs 7 \times 10 \farcs 0$ spectral extraction region at 10 $\mu$m.}
\end{figure}

\clearpage
\begin{figure}[ht]
  \plotone{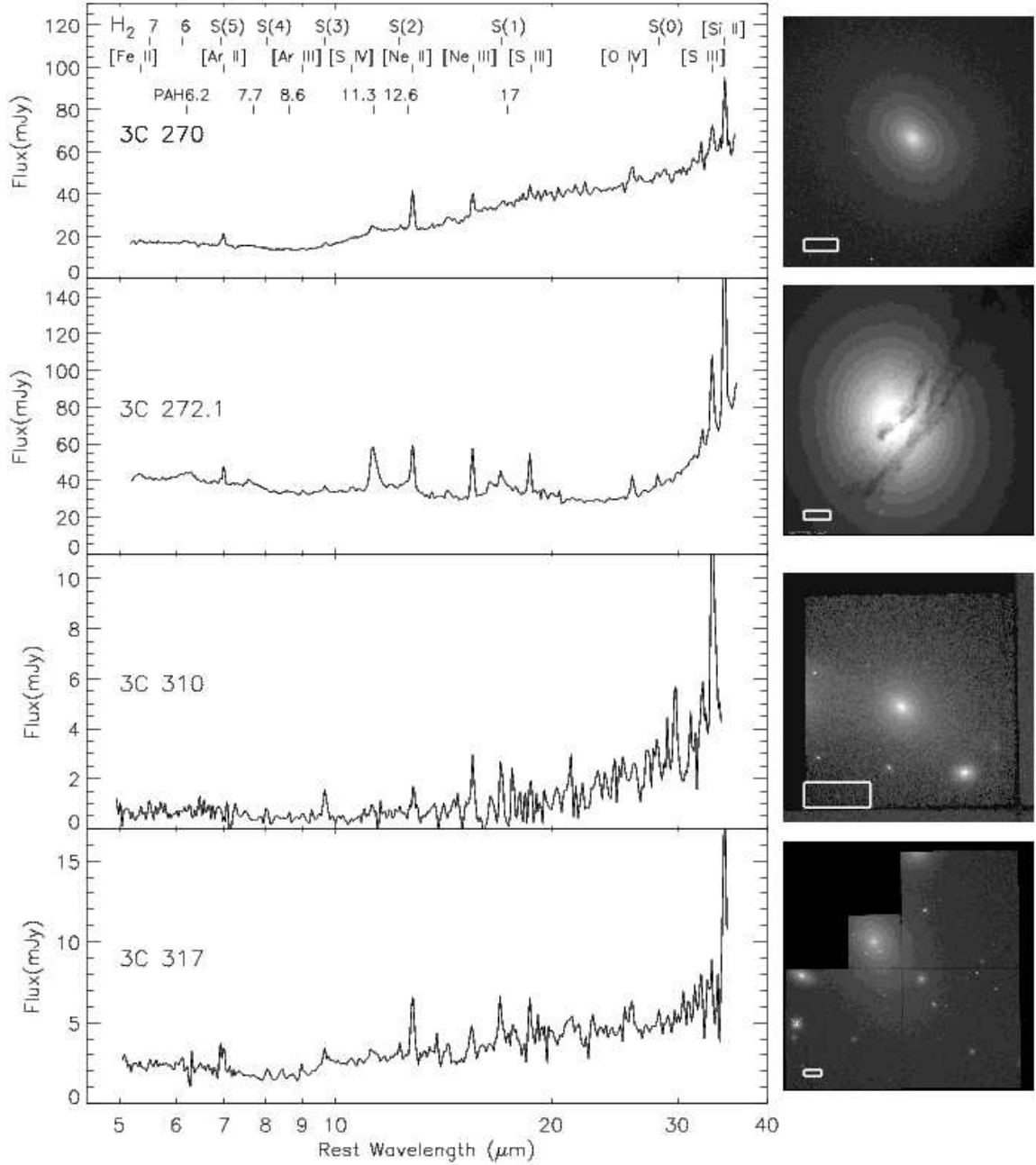}
  \caption{{\it Spitzer} IRS low-resolution spectra of 3C 270 ($=$ NGC 4261), 3C 272.1 ($=$ M 84), 3C 310, and 3C 317. 
              These sources have weak 7.7$\mu$m PAH emission and weak-moderate 9.7 $\mu$m silicate absorption. Note the extremely 
              large PAH 11.3/PAH 7.7 and PAH 17/PAH 7.7 ratios and steeply rising continuum at $>24$ $\mu$m in 3C 272.1. Radio 
              galaxies 3C 270 and 272.1 reside in the Virgo Cluster, 3C 310 in a poor cluster, and 3C 317 in the A 2052 cool-core 
              cluster. The box at the lower left corner of each image shows the size of the 
              $3\farcs 7 \times 10 \farcs 0$ spectral extraction region at 10 $\mu$m.}
\end{figure}

\clearpage
\begin{figure}[ht]
  \plotone{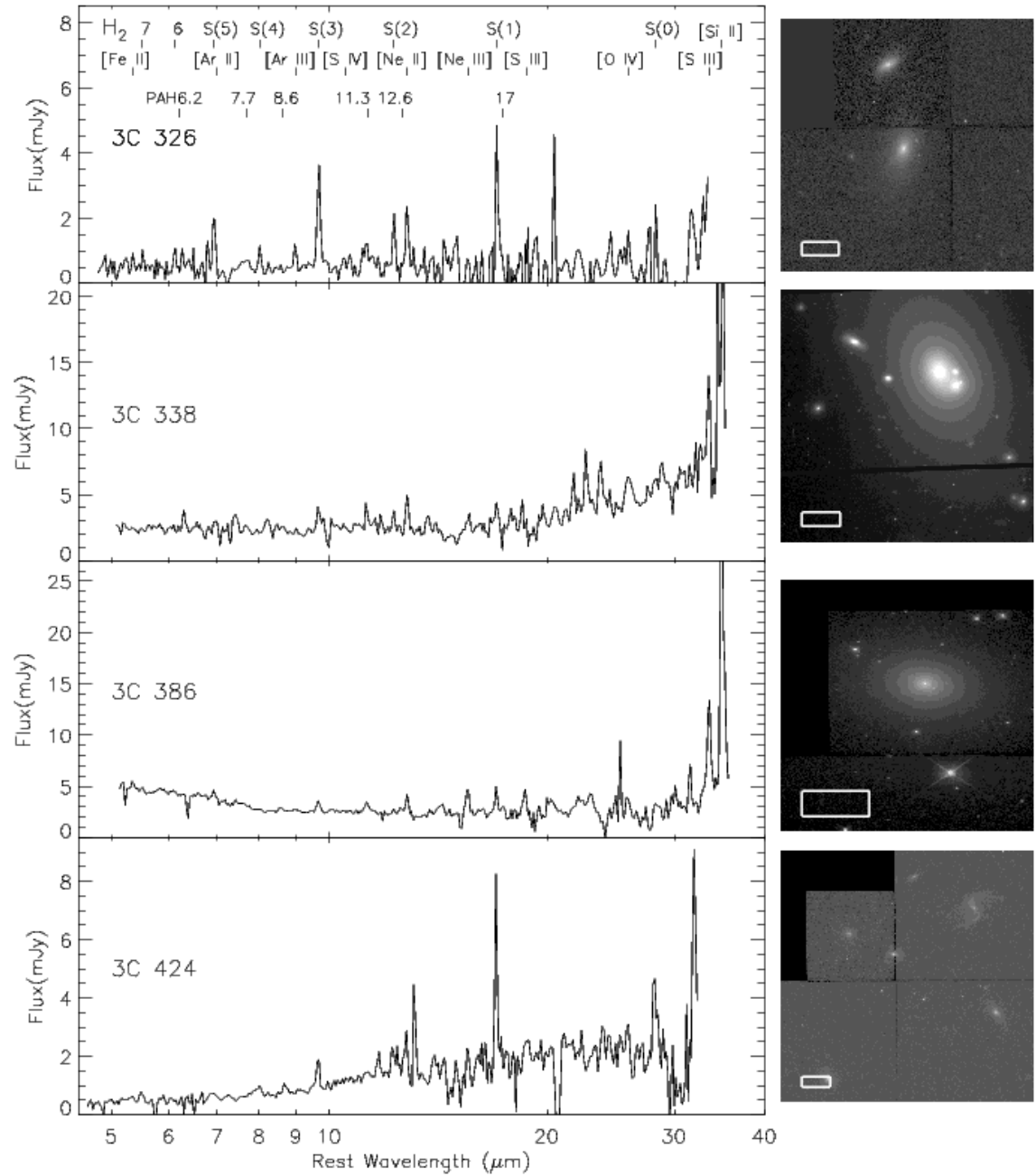}
  \caption{{\it Spitzer} IRS low-resolution spectra of 3C 326, 338, 386, and 424. These sources have weak PAH emission and no 
              silicate absorption. HST WFPC2 and ACS images are to the right. The 3C 326 pair appears to be interacting 
              \citep{oaa07}. Radio galaxy 3C 338 resides in the A 2199 cool core cluster, while 
              3C 386 is an isolated elliptical galaxy. The host of 3C 424 appears (centered on the smaller PC chip) in a rich group of galaxies. 
              The box at the lower left corner of each image shows the size of the 
              $3\farcs 7 \times 10 \farcs 0$ spectral extraction region at 10 $\mu$m.}
\end{figure}

\clearpage
\begin{figure}[ht]
  \plotone{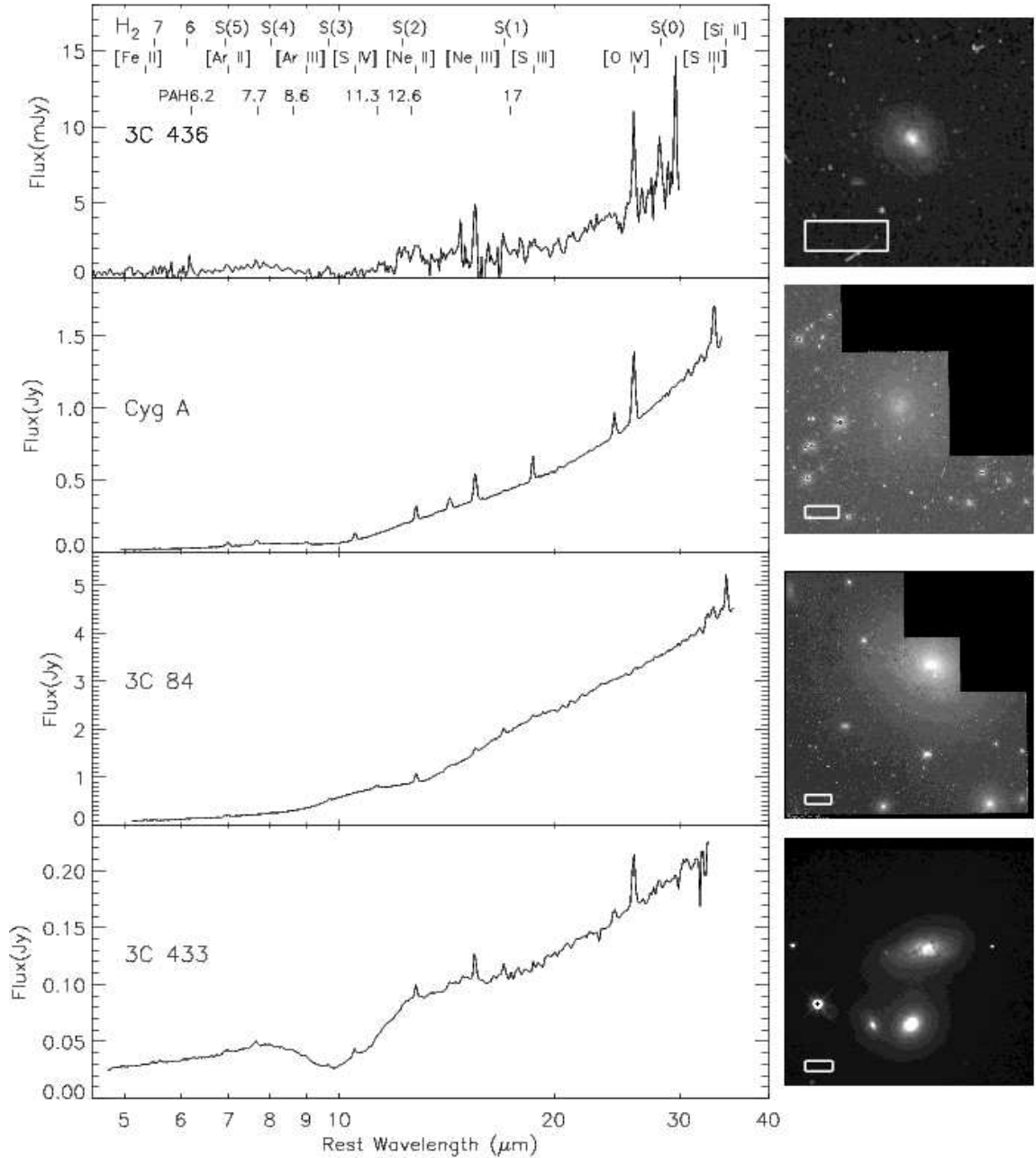}
  \caption{{\it Spitzer} IRS low-resolution spectra of 3C 436, Cyg A, 3C 84 ($=$ Per A), and 3C 433.  
              Silicate emission is seen at 10 $\mu$m and 18 $\mu$m in the Per A spectrum, and silicate absorption in 3C 436, 433, and Cyg A. 
              HST WFPC2 images are at the right. Per A is interacting with an in-falling dusty companion seen in the foreground. 
              The disky, dusty host of 3C 433 resides in an interacting group of 3 galaxies. The box at the lower left corner of each 
              image shows the size of the $3\farcs 7 \times 10 \farcs 0$ spectral extraction region at 10 $\mu$m. See Figs. 5 and 6 
              for spectral fits of 3C 84 and Cyg A.}
\end{figure}

\clearpage
\begin{figure}[ht]
  \plotone{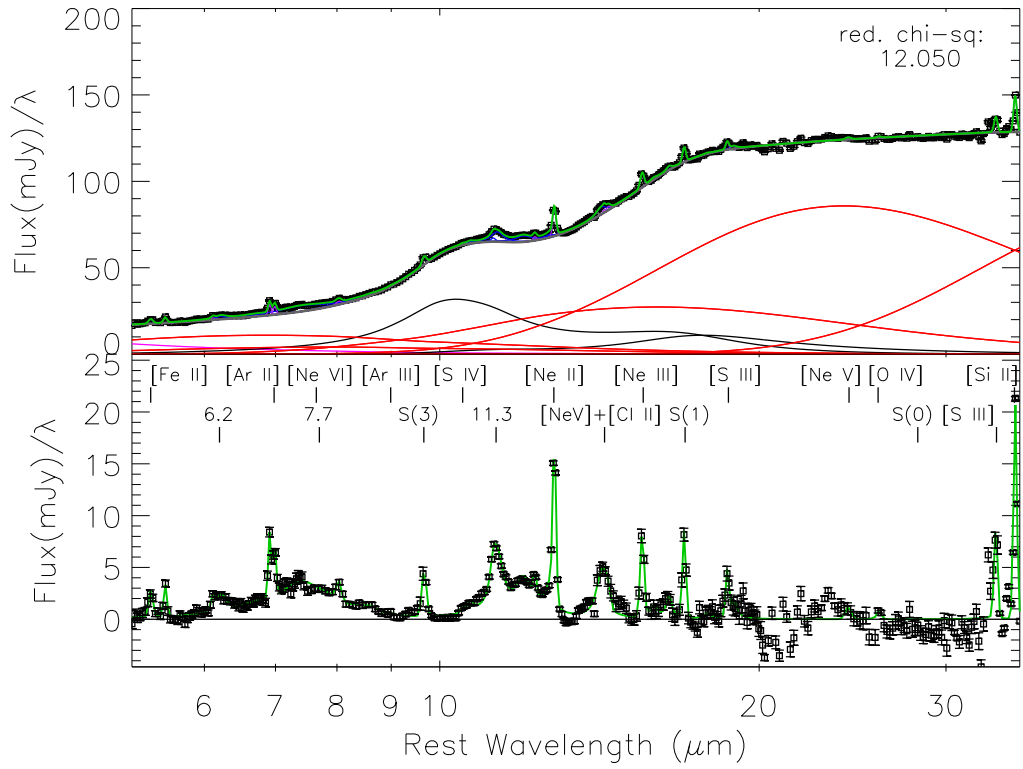}
  \caption{{\it Spitzer} IRS low-resolution spectrum of  3C 84 ($=$ Per A), with best fit model and model components. Silicate 
              emission at 10 $\mu$m and 18 $\mu$m is fit by a blackbody times an empirical AGN silicate dust emissivity curve. 
              The bottom panel shows the continuum-subtracted PAH and emission line spectrum, including H$_2$  pure-rotational 
              lines.}
\end{figure}

\clearpage
\begin{figure}[ht]
  \plotone{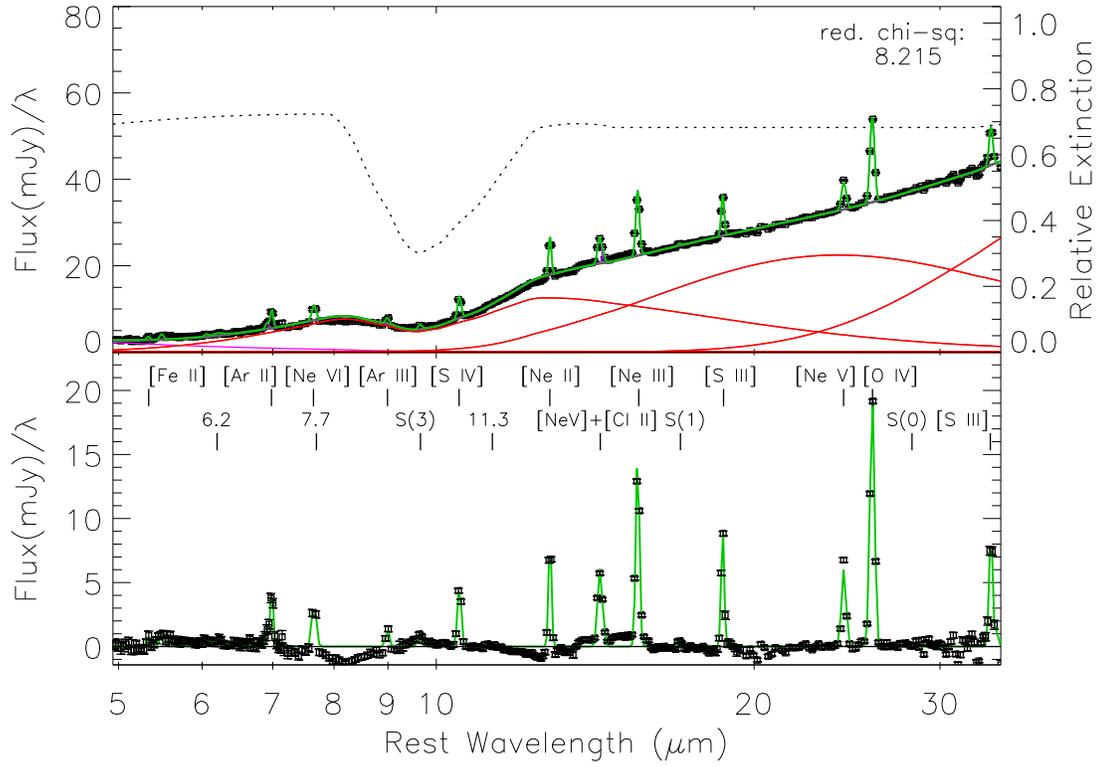}
  \caption{{\it Spitzer} IRS low-resolution spectrum of  3C 405 ($=$ Cyg A), with best fit model and model components.
              A custom extinction curve (dotted line), lacking the 18 $\mu$m silicate feature, was required to fit the continuum.  
              The bottom panel shows the continuum-subtracted emission line spectrum, including H$_2$  pure-rotational 
              lines.}
\end{figure}

\clearpage
\begin{figure}[ht]
  \plotone{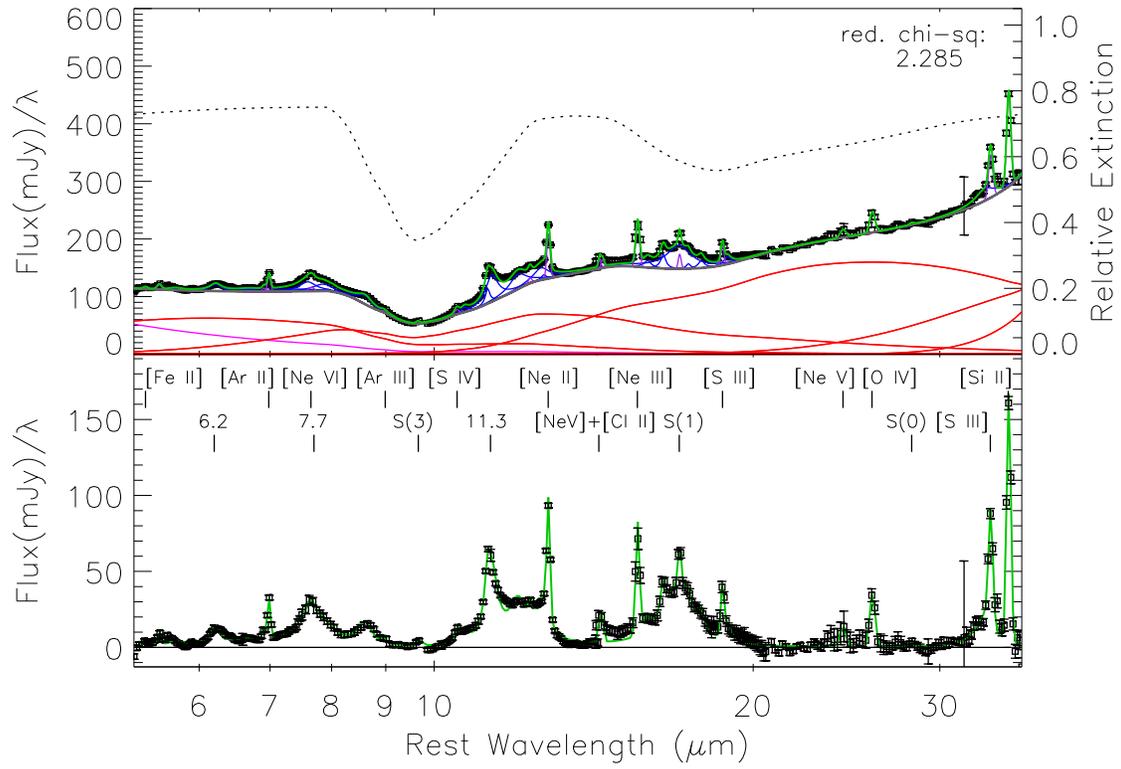}
  \caption{{\it Spitzer} IRS low-resolution spectrum of Cen A ($=$ NGC 5128), with best fit model and model components.              
              The bottom panel shows the continuum-subtracted PAH and emission line spectrum, including H$_2$  pure-rotational 
              lines.}
\end{figure}

\clearpage
\begin{figure}[t]
  \plotone{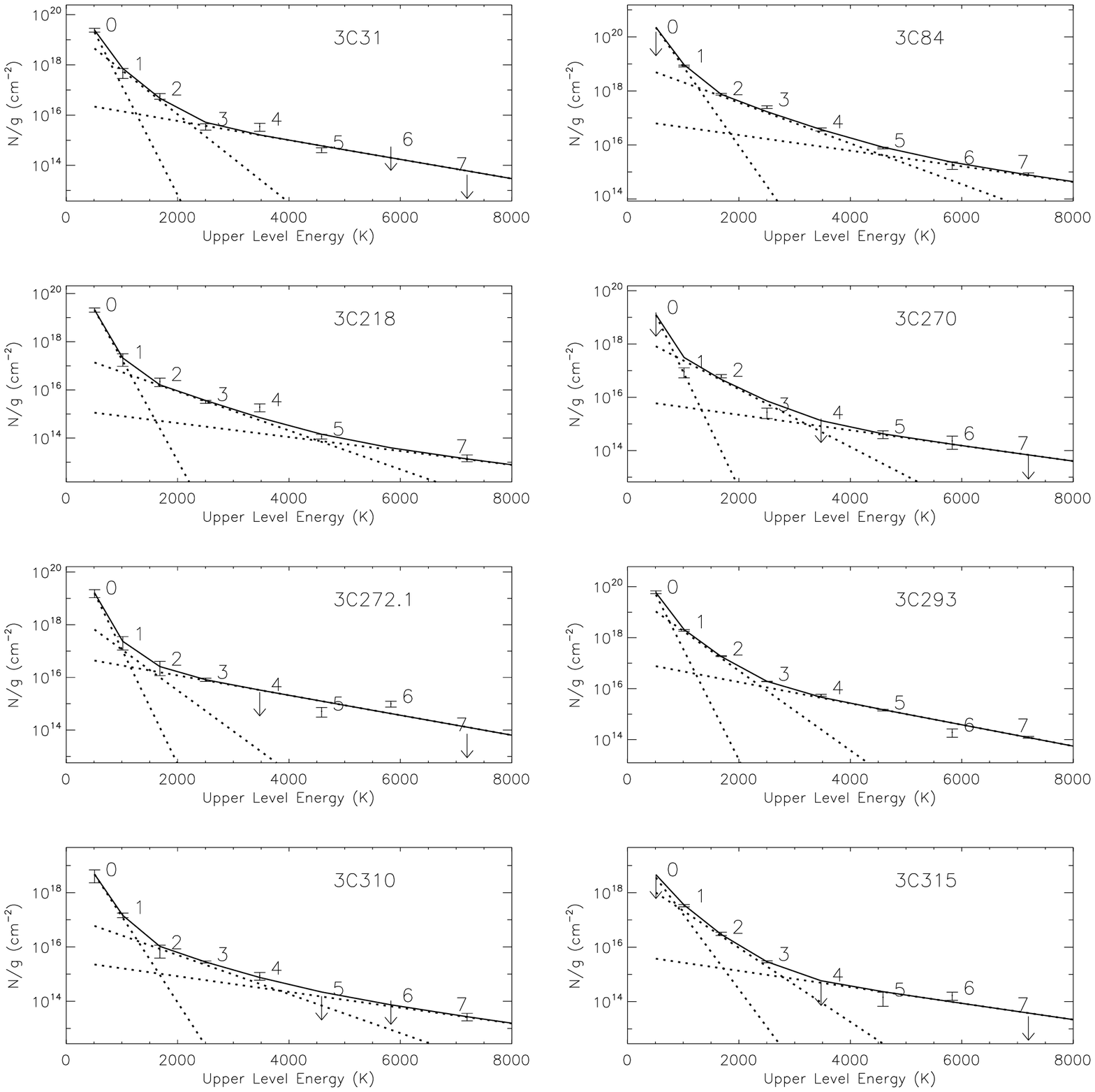}
  \caption{H$_2$ excitation diagrams. Upper level column density divided by statistical weight (for H$_2$ uniformly filling 
              the SL slit width) or 2$\sigma$ upper limit is plotted against upper level energy $E/k_b (K)$ for each 
              H$_2$ 0-0 S(J) pure-rotational emission line.  Model H$_2$ temperature components (dotted lines) are listed in Table 11.}
\end{figure}

\clearpage
\begin{figure}[t]
  \plotone{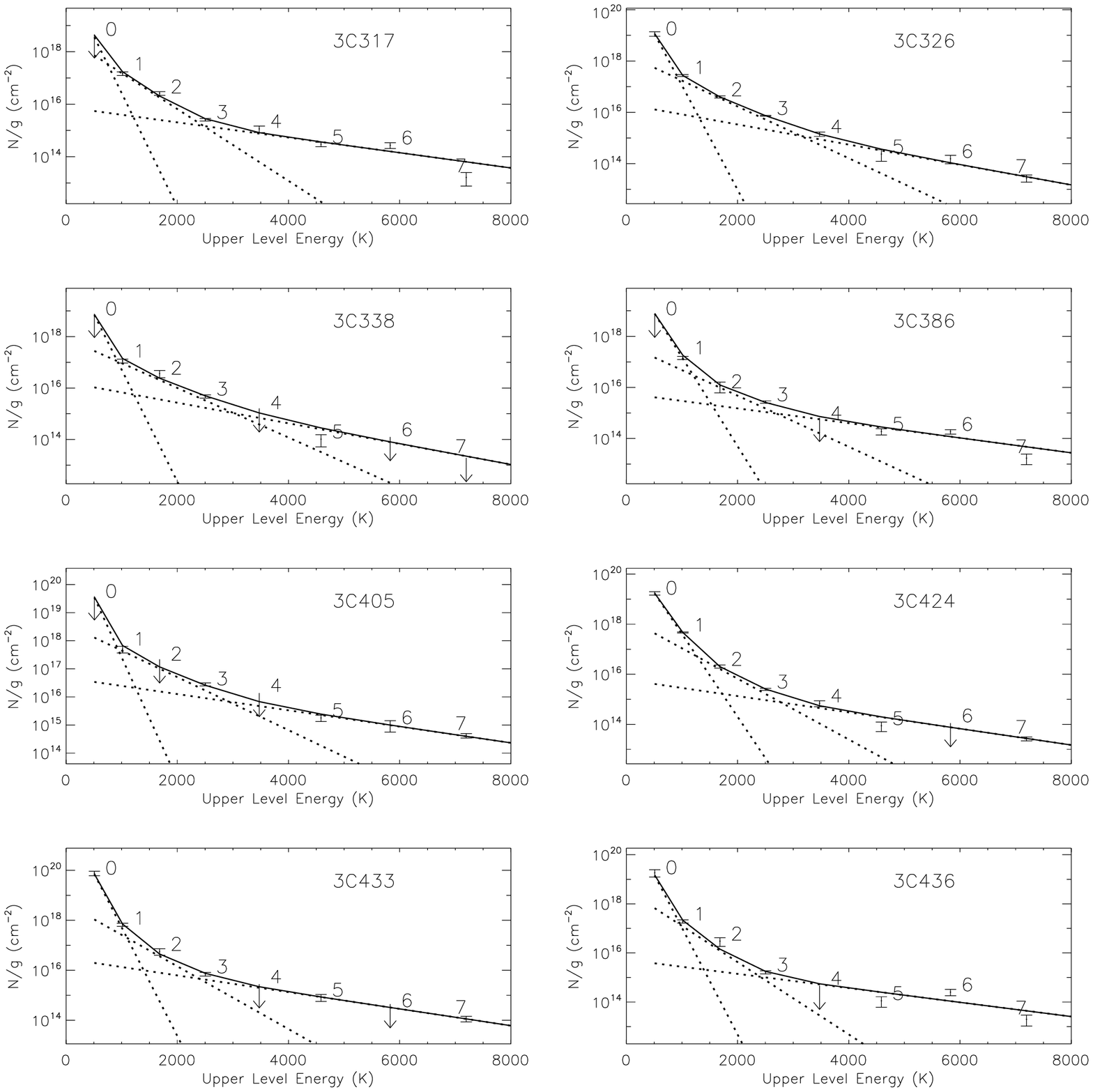}
  \caption{H$_2$ excitation diagrams. Upper level column density divided by statistical weight (for H$_2$ uniformly filling 
              the SL slit width) or 2$\sigma$ upper limit is plotted against upper level energy $E/k_b (K)$ for each 
              H$_2$ 0-0 S(J) pure-rotational emission line.  Model H$_2$ temperature components (dotted lines) are listed in Table 11.}
\end{figure}

\clearpage
\begin{figure}[t]
  \plotone{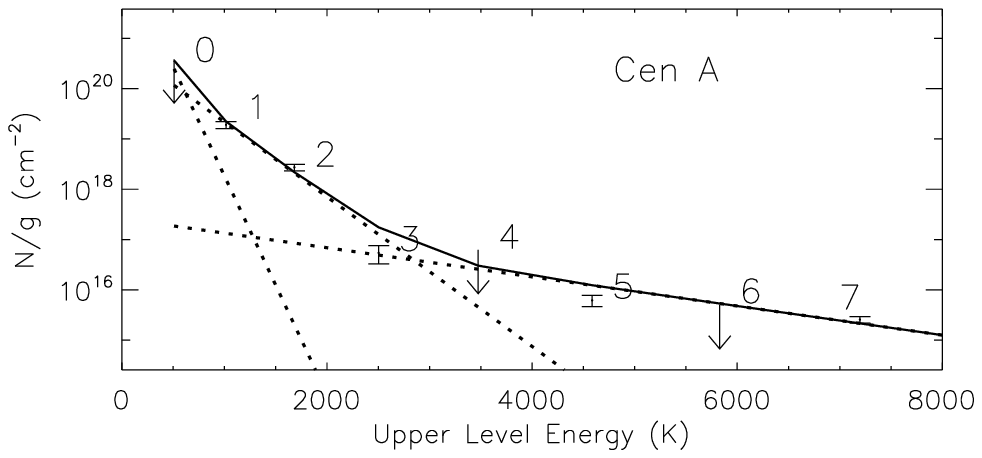}
  \caption{H$_2$ excitation diagram for Cen A. Upper level column density divided by statistical weight (for H$_2$ uniformly filling 
              the SL slit width) or 2$\sigma$ upper limit is plotted against upper level energy $E/k_b (K)$ for each 
              H$_2$ 0-0 S(J) pure-rotational emission line.  Model H$_2$ temperature components (dotted lines) are listed in Table 11.}
\end{figure}

\clearpage
\begin{figure}[ht]
   \plotone{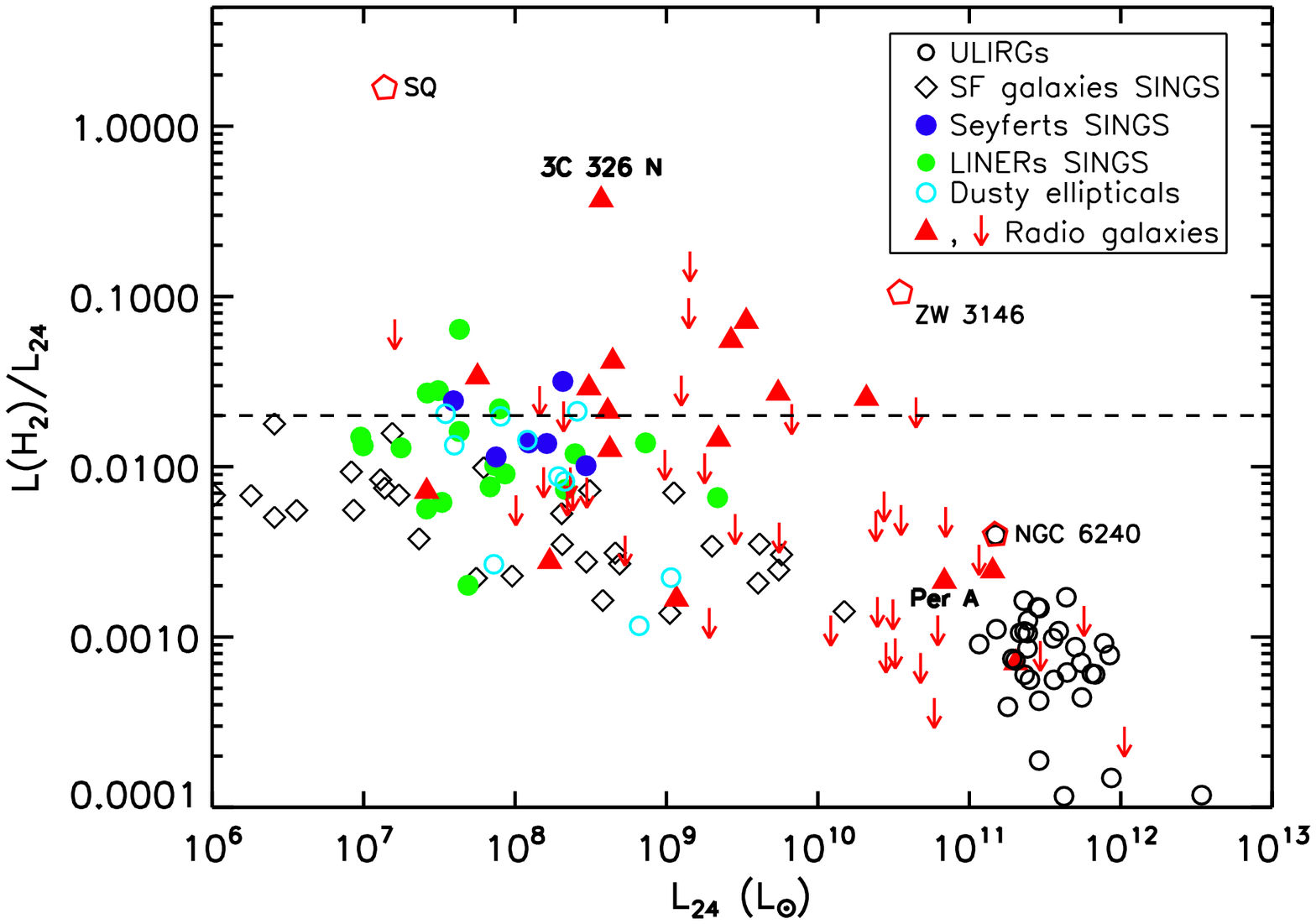}
   \caption{Ratio of H$_2$ luminosity summed over the 0-0 S(0)-S(3) pure rotational lines to $L_{24}=\nu L_\nu$(24 $\mu$m, rest) 
               luminosity. Radio galaxies (this work) are compared to SINGS star-forming galaxies, Seyferts, and LINERs \citep{rhh07}, 
               ULIRGs \citep{hah06}, and the subset of \cite{kos08} dusty ellipticals that have 24 $\mu$m {\it Spitzer} MIPS fluxes \citep{tbm07}. 
               Radio galaxy H$_2$ upper limits (from Tables 2 and 6) are plotted as downward pointing arrows.
               The LIRG NGC 6240 \citep{abs06}, the Zw 3146 brightest cluster galaxy \citep{e06}, and the Stephan's Quintet intergalactic shock 
               \citep[SQ shock sub-region, ][]{ca10} are plotted for comparison. The large values of
               $L$(H$_2$)/$L_\mathrm{24}>0.02$ in many radio galaxies indicate the importance of non-radiative heating (e.g., shocks).}
\end{figure}

\clearpage
\begin{figure}[ht]
   \plotone{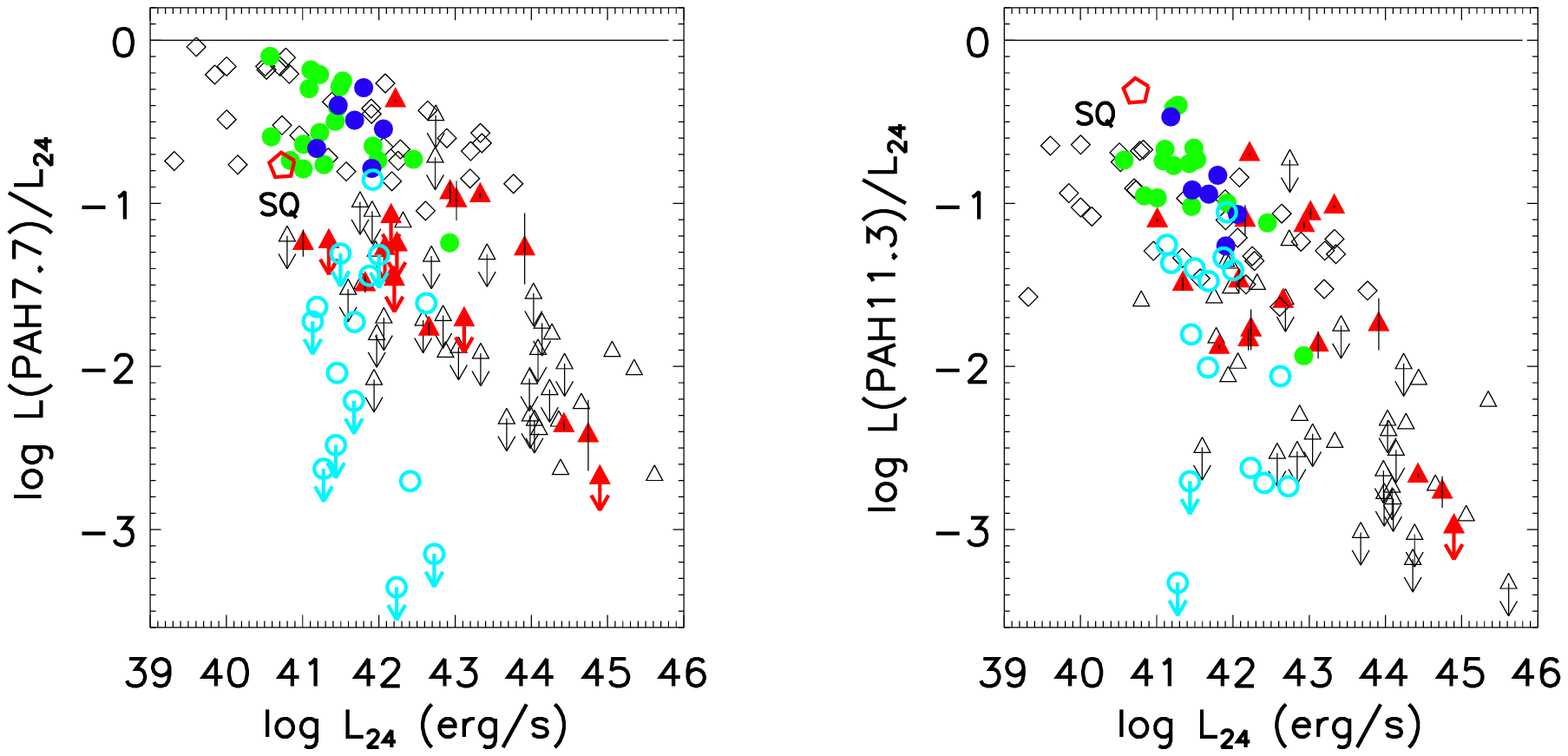}
   \caption{PAH to $\nu L_\nu$(24 $\mu$m, rest) luminosity ratio, for the 7.7 and 11.3 $\mu$m PAH features. (See Fig. 11 for symbol legend; H$_2$ 
               nondetected radio galaxies are plotted as open triangles.) Solid line corresponds to 1:1 ratio.  Radio galaxies (this work) are compared 
               to SINGS star-forming galaxies, Seyferts, and LINERs \citep{rhh07,sdd07}, the \cite{kos08} dusty ellipticals, and the Stephan's Quintet 
               intergalactic shock \citep[SQ shock sub-region, ][]{ca10}. Most radio galaxies have $L(\mathrm{PAH}7.7)/L_{24}<0.1$, indicating a very 
               small star formation contribution and dominant AGN contribution to the mid-IR continuum. }
\end{figure}

\begin{figure}[ht]
   \plotone{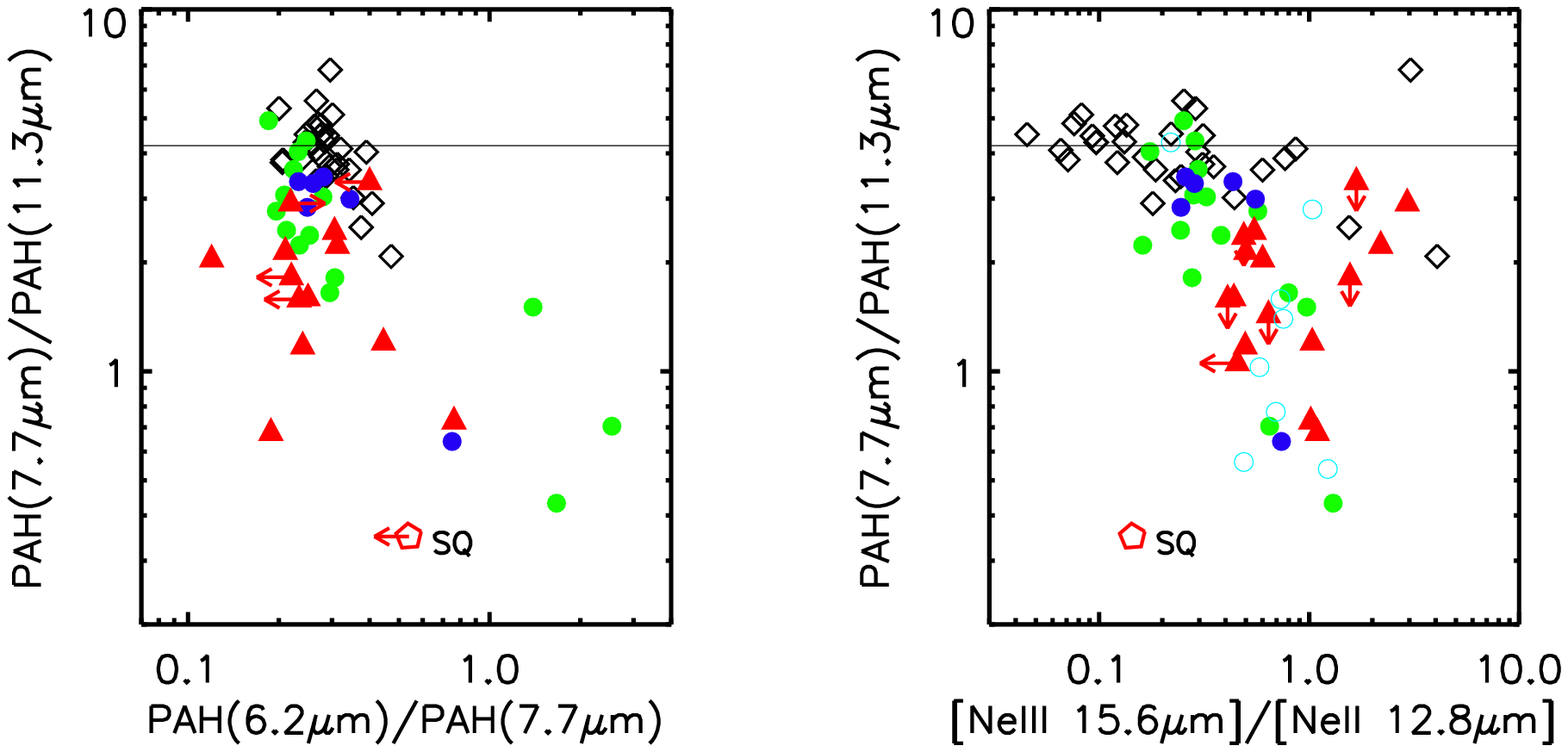}
   \caption{Left: PAH emission feature flux ratios (see Fig. 11 for symbol legend).  H$_2$-detected radio galaxies (triangles, this work) are compared to
               SINGS star-forming galaxies, Seyferts, and LINERs \citep{sdd07}, the \cite{kos08} dusty ellipticals, 
               and the Stephan's Quintet intergalactic shock \citep[SQ shock sub-region, ][]{ca10}. Four radio galaxies with 
               neither 6.2 $\mu$m nor 7.7 $\mu$m PAH detected are not plotted. The median
               PAH(7.7 $\mu$m)/PAH(11.3 $\mu$m) ratio is lower in radio galaxies than in star-forming galaxies,
               while the median PAH(6.2 $\mu$m)/PAH(7.7 $\mu$m) ratio is similar. This indicates lower PAH ionization caused 
               by a weaker interstellar UV radiation field or increased gas-phase ionization by X-rays or cosmic rays.
               Right: PAH 7.7 $\mu$m/PAH 11.3 $\mu$m vs. [Ne {\sc iii}]/[Ne {\sc ii}] line flux ratio after \cite{sdd07}.
               One radio galaxy with no PAH features detected (3C 405) is not plotted. The large
               [Ne {\sc iii}]/[Ne {\sc ii}] ratio relative to star-forming galaxies may be attributed to AGN activity.}
\end{figure}

\clearpage
\begin{figure}[ht]
   \plotone{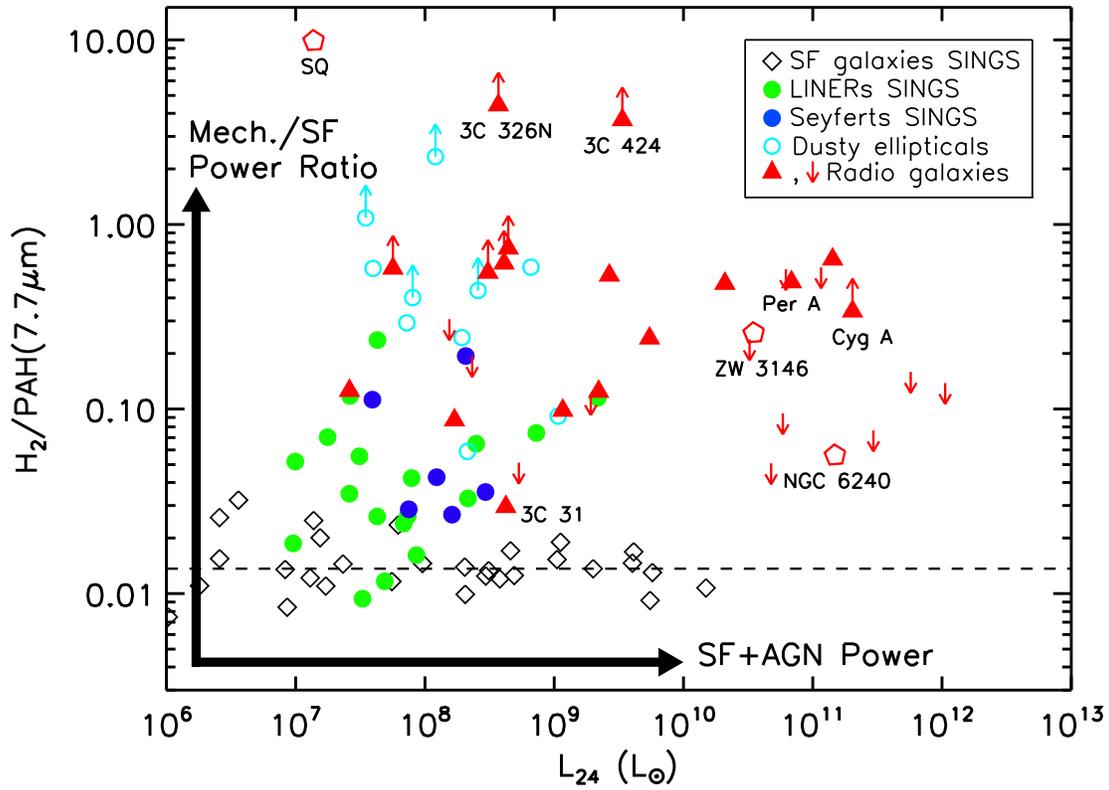}
   \caption{Ratio of H$_2$ luminosity summed over the 0-0 S(0)-S(3) lines to 7.7 $\mu$m PAH luminosity vs.  
               $\nu L_\nu$(24 $\mu$m,rest) continuum luminosity. This ratio indicates the relative importance of mechanical 
               heating and star formation power. All but one of the H$_2$-detected radio galaxies from our sample 
               stand out above normal star-forming galaxies from the SINGS survey \citep{rhh07} and are MOHEGs, 
               with $L$(H$_2$)/$L(\mathrm{PAH 7.7})>0.04$. Radio galaxy H$_2$ upper limits with detected 7.7 $\mu$m PAH emission
               (from Tables 2 and 6) are plotted as downward pointing arrows. Many of the LINERs and Seyferts from the 
               SINGS sample \citep{rhh07} and several of the \cite{kos08} dusty ellipticals are MOHEGs. Other MOHEGs from the 
               literature, including NGC 6240 \citep{abs06}, Zw 3146 \citep{e06}, and the Stephan's Quintet intergalactic shock 
               \citep[SQ shock sub-region, ][]{ca10}, are plotted for comparison.}
\end{figure}

\clearpage
\begin{figure}[ht]
   \plotone{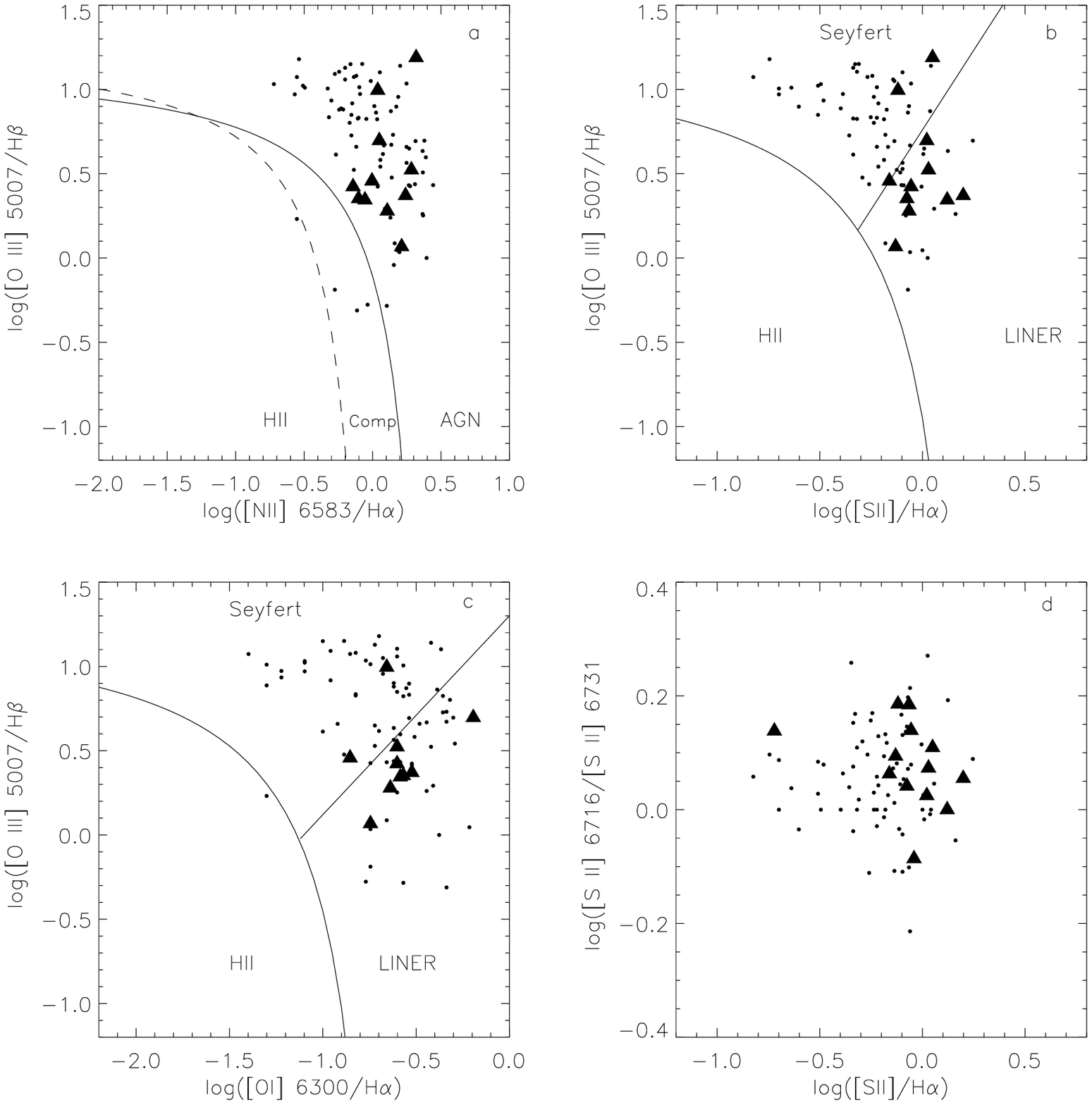}
   \caption{BPT \citep{bpt81} AGN diagnostic diagrams. a) The radio MOHEGs (filled triangles) have optical emission line ratios consistent
               with pure AGN, rather than star-forming (H {\sc ii}) or composite galaxies. Optical line fluxes for these galaxies and other $z<0.3$ 3C 
               radio galaxies (small dots) are from  \cite{bcc09}. b) and c) Nearly all radio MOHEGs on these plots (except 3C 433 and 436) are 
               classified as LINERs according to the dividing lines of \cite{kgk06}.  d) The density-sensitive [S {\sc ii}] 
               doublet ratio falls within the normal range for radio galaxies.}
\end{figure}

\clearpage
\begin{figure}[ht]
   \plotone{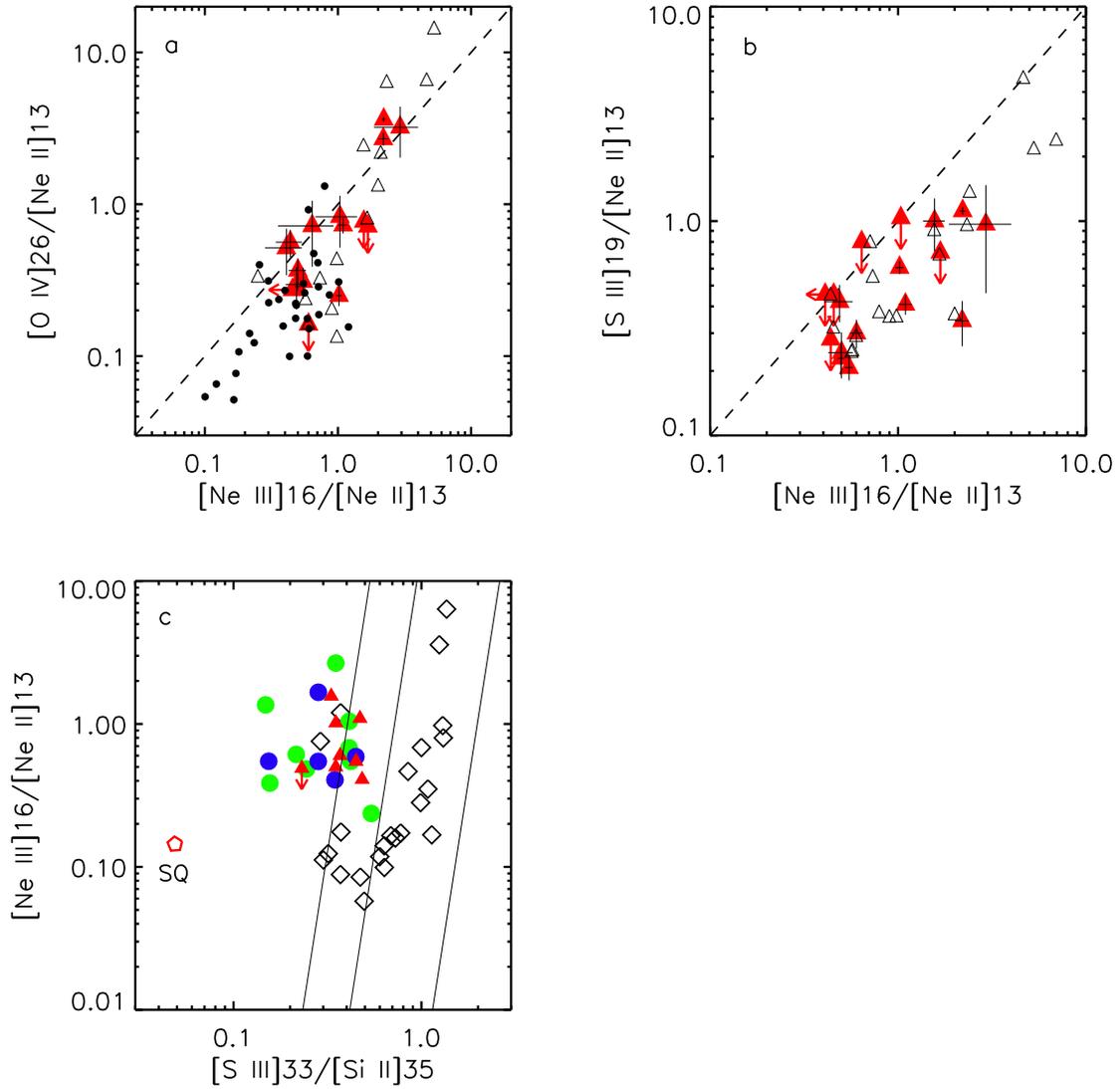}
   \caption{a,b) The mid-IR forbidden line ratios of radio MOHEGs (filled triangles) and H$_2$ undetected radio galaxies (open triangles).
               The dashed lines correspond to a 1:1 ratio of ratios. Other LINERs measured by {\it Spitzer} \citep[small dots]{dsm09} are shown for 
               comparison in the first panel. c) \cite{dsa06} MIR forbidden line diagnostic diagram (see Fig. 14 for symbol legend). The line ratios 
               of radio MOHEGs with [Si {\sc ii}] measured are consistent with LINERs and Seyferts, but segregate 
               from star-forming galaxies in the SINGS survey \citep{sdd07}. The Stephan's Quintet intergalactic shock
               \citep[red pentagon, SQ shock sub-region, ][]{ca10} has lower ionization than radio MOHEGs.}
\end{figure}

\clearpage
\begin{figure}[ht]
   \plotone{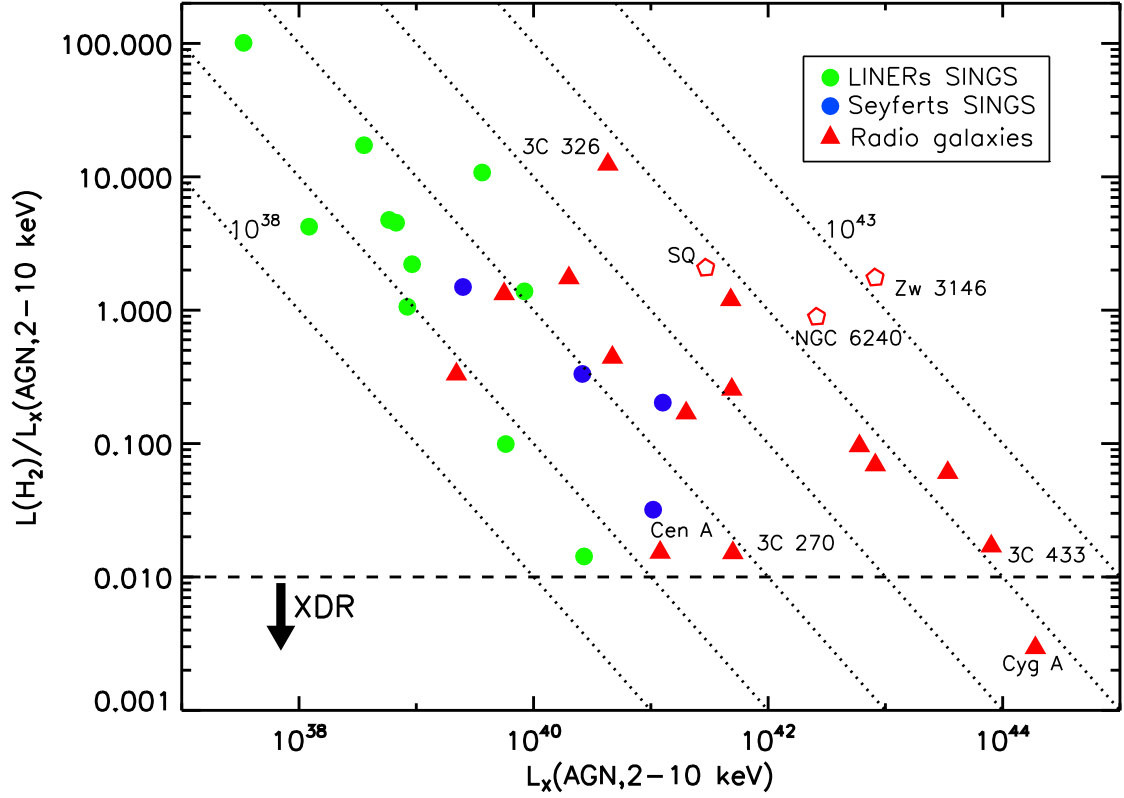}
   \caption{Ratio of H$_2$ luminosity (summed over 0-0 S(0)-S(3)) to unabsorbed AGN X-ray (2-10 keV) luminosity. For
               Stephan's Quintet \citep[SQ main shock, ][]{ca10}, the H$_2$ and X-ray luminosities (0.001-10 keV) 
               are integrated over the entire main shock region. For NGC 6240, the X-ray luminosity (0.1-10 keV) is the total for the
               binary AGN \citep{kbh03}. There is likely significant contamination to the X-ray luminosity of Zw 3146 from 
               the surrounding galaxy cluster. Two radio galaxies (3C 310 and 424) that have not been observed by {\it Chandra} 
               are not plotted. Diagonal dotted lines indicate constant $L$(H$_2$). The horizontal dashed 
               line indicates the 1\% maximum H$_2$ to X-ray luminosity ratio for an X-ray dissociation region (XDR)
               at the characteristic observed H$_2$ temperature of 200 K. The H$_2$ emission from all radio MOHEGs (except
               Cyg A) and SINGS AGNs is too luminous to be powered by X-rays from the active nuclei. }
\end{figure}

\clearpage
\begin{figure}[ht]
   \plotone{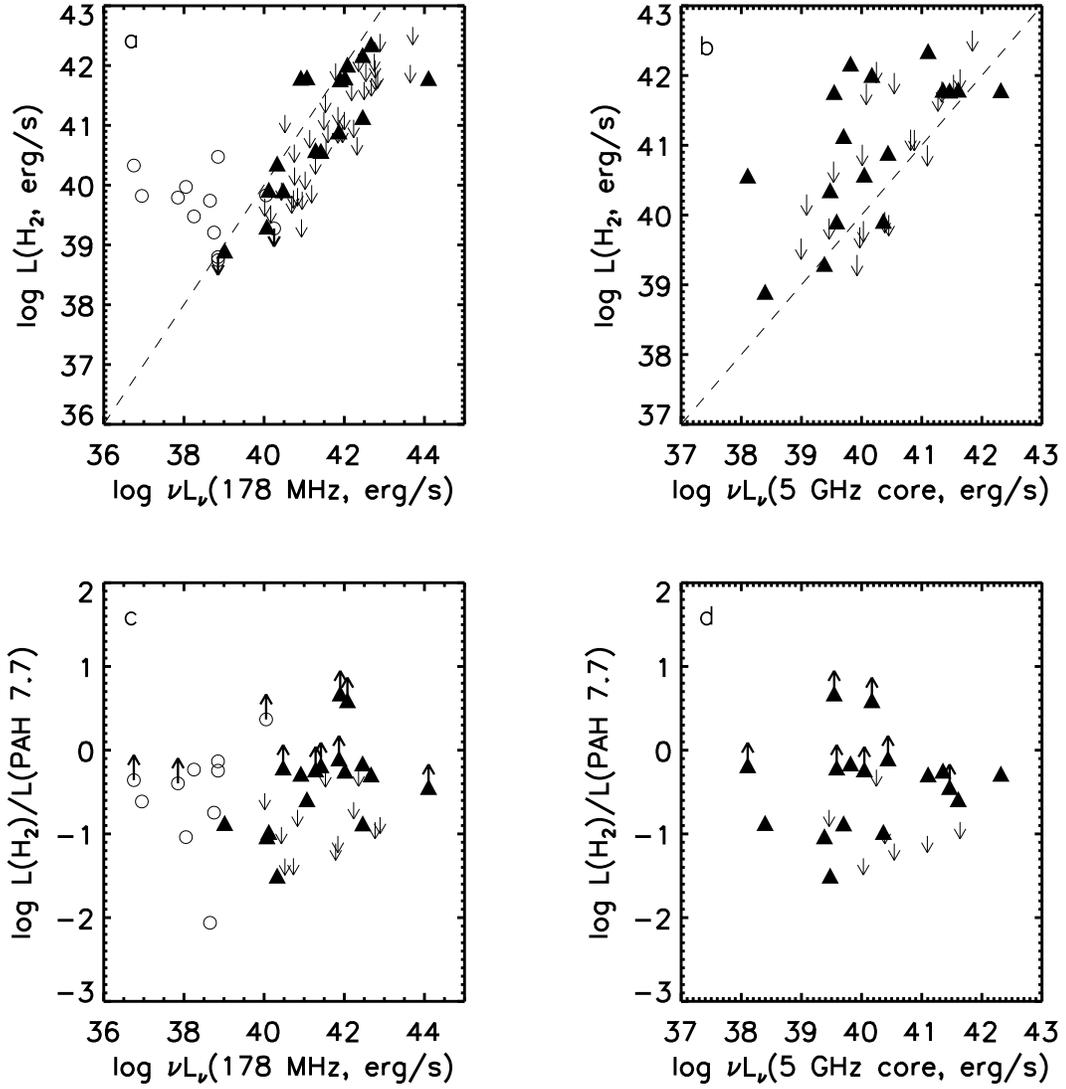}
   \caption{a,b) H$_2$ luminosity (summed over 0-0 S(0)-S(3)) vs. radio luminosity at 178 MHz and arcsecond-scale radio core 
               luminosity at 5 GHz (http://3crr.extragalactic.info/cgi/database). 
               Radio MOHEGs (filled triangles), H$_2$ nondetected radio galaxies (downward arrows), and \cite{kos08} dusty ellipticals 
               (open circles) are shown. Dashed lines correspond to 1:1 ratio. 
               c,d) H$_2$ over 7.7 $\mu$m PAH luminosity ratio vs. radio luminosity at 178 MHz and arcsecond-scale radio core luminosity 
               at 5 GHz.}
\end{figure}

\clearpage

\begin{deluxetable}{llllrrc}
\tablecaption{H$_2$ Detected Radio Galaxies (Radio MOHEGs)}
\tablehead{
\colhead{Source} & \colhead{Morph.\tablenotemark{a}} & \colhead{Sp.T. \tablenotemark{b}} & \colhead{z}  & \colhead{~~$F_{24}$\tablenotemark{c}} & \colhead{~~S178\tablenotemark{d}} & \colhead{Ref.\tablenotemark{e}}
}

\startdata
3C 31    & I,TJ   & LIG & 0.01701  &   19.9$\pm$0.3    &   18.3 & 1 \\  
3C 84    & I,FD   & LIG & 0.01756  &  3020.$\pm$30.    &   66.8 & 1 \\  
3C 218   & I,TJ   & LIG & 0.05488  &    9.1$\pm$0.2    &  225.7 & 4 \\  
3C 270   & I,TJ   & LIG & 0.007465 &   42.4$\pm$0.4    &   53.4 & 2 \\  
3C 272.1 & I,TJ   & LIG & 0.003536 &   29.4$\pm$0.2    &   21.1 & 1 \\  
3C 293   & I,CSC  & LIG & 0.04503  &   34.3$\pm$1.0    &   13.8 & 1 \\  
3C 317   & I,FD   & LIG & 0.03446  &    4.5$\pm$0.1    &   53.4 & 3 \\  
3C 338   & Ip,FD  & LIG & 0.03035  &    4.4$\pm$0.3    &   51.1 & 3 \\  
3C 386   & I,FD   & LIG & 0.01689  &    2.7$\pm$0.4    &   26.1 & 1 \\  
3C 424   & I,DD   & LIG & 0.1270   &    2.2$\pm$0.1    &   15.9 & 3 \\  
Cen A    & I,TJ   & LIG & 0.001825 &  4950.$\pm$110.   & 1005.  & 4 \\  
\\
3C 310   & II,FD  & LIG & 0.0538   &    1.9$\pm$0.2    &   60.1 & 1 \\  
3C 315   & II,X,CSC & HIG & 0.1083 &    2.5$\pm$0.1    &   19.4 & 1 \\  
3C 326 N & II     & LIG & 0.0895   &   0.53$\pm$0.12   &   22.2 & 1 \\  
3C 405   & II     & HIG & 0.05608  &   800.$\pm$20.    & 9483.  & 3 \\  
3C 433   & II,X   & HIG & 0.1016   &   154.$\pm$ 3.    &   61.3 & 1 \\  
3C 436   & II     & HIG & 0.2145   &    4.0$\pm$0.1    &   19.4 & 1 \\  
\enddata
\tablenotetext{a}{Radio morphologies: I or II $=$ \cite{fr74} type (p $=$ peculiar), TJ $=$ twin jet, 
                  FD $=$ fat double, CSC $=$ compact symmetric core, DD $=$ double-double, X $=$ 'X'-shaped, SJ $=$ single (one-sided) jet.}
\tablenotetext{b}{Optical spectral type: LIG $=$ low ionization narrow line galaxy, HIG $=$ high ionization narrow line galaxy,  BLRG $=$ broad line radio 
                  galaxy.}
\tablenotetext{c}{Rest 24 $\mu$m flux (mJy) measured with {\it Spitzer} IRS. A 3.7$\sigma$-clipped average over 22.5-25.5 $\mu$m was used, removing
                  any large noise spikes or contributions from the [Ne {\sc v}] 24.31 $\mu$m line.}
\tablenotetext{d}{The 178 MHz flux density (Jy) on \cite{bgp77} scale.}
\tablenotetext{e}{References for 178 MHz flux density. (1) \cite{lrl83} (2) \cite{kwp81} (3) \cite{kpw69} (4) Estimated from \cite{kwp81} 160 MHz flux density.}
\end{deluxetable}

\clearpage
\begin{deluxetable}{llllrrc}
\tablecaption{H$_2$ Nondetected Radio Galaxies}
\tablehead{
\colhead{Source} & \colhead{Morph.\tablenotemark{a}} & \colhead{Sp.T.\tablenotemark{b}} & \colhead{z}  & \colhead{~~$F_{24}$\tablenotemark{c}} & \colhead{~~S178\tablenotemark{d}} & \colhead{Ref.\tablenotemark{e}} 
}

\startdata
3C 15    & I     & LIG  & 0.0730   &   12.4$\pm$0.3  &   16.9 & 2 \\   
3C 29    & I,TJ  & LIG  & 0.045031 &   11.2$\pm$0.3  &   16.2 & 2 \\   
3C 66B   & I,TJ  & LIG  & 0.02126  &    6.6$\pm$0.1  &   26.8 & 1 \\   
3C 76.1  & I,FD  & LIG  & 0.03249  &    2.6$\pm$0.2  &   13.3 & 1 \\   
3C 83.1  & I     & LIG  & 0.02514  &    4.9$\pm$0.1  &   26.6 & 1 \\   
3C 120   & I,SJ  & BLRG & 0.03301  &   573.$\pm$5.   &    7.4 & 2 \\   
3C 129   & I,TJ  & LIG  & 0.0208   &    7.5$\pm$0.2  &   51.1 & 3 \\   
3C 189   & I,TJ  & LIG  & 0.04284  &    6.8$\pm$0.1  &    7.4 & 2 \\   
3C 264   & I,TJ  & LIG  & 0.02172  &   15.2$\pm$0.3  &   28.3 & 1 \\   
3C 274   & I,TJ  & LIG  & 0.004360 &    75.$\pm$1.   & 1145.  & 1 \\   
3C 318.1 & I     & LIG  & 0.04531  &    0.9$\pm$0.2  &   12.3 & 3 \\   
3C 348   & I,TJ  & LIG  & 0.1540   &    0.6$\pm$0.2  &  387.  & 2 \\   
3C 403.1 & I     & LIG  & 0.0554   &       $<0.2$    &   14.7 & 3 \\   
3C 449   & I,TJ  & LIG  & 0.017085 &   0.75$\pm$0.11 &   12.5 & 1 \\   
3C 465   & I,TJ  & LIG  & 0.03022  &    4.3$\pm$0.2  &   41.2 & 1 \\   
NGC 6251 & Ip    & LIG  & 0.02471  &    42.$\pm$2.   &   10.9 & 1 \\   
IC 4296  & Ip,DD & LIG  & 0.01247  &   13.7$\pm$0.2  &   16.8 & 2 \\   
 \\                                   
3C 17    & IIp   & BLRG & 0.219685 &  12.6$\pm$0.4   &  22.2 & 2 \\    
3C 28    & II    & LIG  & 0.1953   &      $<0.3$     &  17.8 & 1 \\    
3C 33    & II    & HIG  & 0.0597   &  97.1$\pm$0.8   &  59.3 & 1 \\    
3C 61.1  & II    & HIG  & 0.1878   &   7.2$\pm$0.3   &  34.0 & 1 \\    
3C 111   & II    & BLRG & 0.0485   &  168.$\pm$1.    &  70.4 & 3 \\    
3C 123   & IIp   & LIG  & 0.2177   &   8.2$\pm$0.2   & 206.0 & 1 \\    
3C 192   & II    & HIG  & 0.0597   &   9.8$\pm$0.4   &  23.0 & 1 \\    
3C 219   & II    & BLRG & 0.1744   &  11.1$\pm$0.1   &  44.9 & 1 \\    
3C 234   & II    & HIG  & 0.1848   &  289.$\pm$1.    &  34.2 & 1 \\    
3C 319   & II    & LIG  & 0.1920   &  0.35$\pm$0.10  &  16.7 & 1 \\    
3C 321   & II    & HIG  & 0.0961   &  360.$\pm$10.   &  14.7 & 1 \\    
3C 346   & II,FD & HIG  & 0.1620   &   9.0$\pm$0.1   &  11.9 & 1 \\    
3C 381   & II    & HIG  & 0.1605   &  44.2$\pm$0.5   &  18.1 & 1 \\    
3C 382   & II    & BLRG & 0.0579   &   91.$\pm$2.    &  21.7 & 1 \\    
3C 388   & II    & LIG  & 0.0917   &   1.7$\pm$0.1   &  26.8 & 1 \\    
3C 390.3 & II    & BLRG & 0.0561   &  231.$\pm$1.    &  51.8 & 1 \\    
3C 401   & II    & LIG  & 0.2011   &   1.5$\pm$0.1   &  22.8 & 1 \\    
3C 445   & II    & BLRG & 0.0562   &  243.$\pm$2.    &  25.2 & 2 \\    
3C 452   & II    & HIG  & 0.0811   &  57.6$\pm$0.3   &  59.3 & 1 \\    
3C 459   & II    & HIG  & 0.2199   &  103.$\pm$3.    &  30.8 & 2 \\    
Pic A    & II    & BLRG & 0.03506  &  129.$\pm$2.    & 411.  & 4 \\    
\enddata
\tablenotetext{a}{Radio morphology (see Table 1 for key).}
\tablenotetext{b}{Optical spectral type (see Table 1 for key).}
\tablenotetext{c}{Rest 24 $\mu$m flux (mJy) or 2$\sigma$ upper limit measured with {\it Spitzer} IRS.  A 3.7$\sigma$-clipped average over 22.5-25.5 $\mu$m
                  was used, removing any large noise spikes or contributions from the [Ne {\sc v}] 24.31 $\mu$m line.}
\tablenotetext{d}{The 178 MHz flux density (Jy) on \cite{bgp77} scale.} 
\tablenotetext{e}{References for 178 MHz flux density. (1) \cite{lrl83}. (2) \cite{kwp81} (3) \cite{kpw69} (4) Estimated from \cite{kwp81} 160 MHz flux density.}
\end{deluxetable}

\clearpage
\begin{deluxetable}{lllllll}
\tablecaption{Radio MOHEG Host, Environment, and Interactions}
\tablehead{
\colhead{Source} & \colhead{Name} & \colhead{NGC} & \colhead{Host\tablenotemark{a}}  & \colhead{Dust Morph.\tablenotemark{b}} & \colhead{Environment\tablenotemark{c}} & \colhead{Interaction\tablenotemark{d}} 
}

\startdata                                                                                   
3C 31    &       &  383   & Ep   & disk      & Arp 331 group      & close companion  \\      
3C 84    & Per A & 1275   & cD+D & complex   & Perseus CC cluster & infalling spiral \\      
3C 218   & Hyd A &        & cD   &           & A 780 CC cluster   & close companion  \\      
3C 270   &       & 4261   & Ep   & disk      & Virgo cluster      & tail             \\      
3C 272.1 & M 84  & 4374   & Ep   & lanes     & Virgo cluster      &                  \\      
3C 293   &       &        & S0   & lanes     & pair               & bridge,tail      \\      
3C 310   &       &        & Ep+D &           & poor  cluster      & bridge           \\      
3C 315   &       &        & S0   & lane      & pair               & close companion  \\      
3C 317   &       &        & cD+D & filaments & A2052 CC cluster   & close companion  \\      
3C 326 N &       &        & Ep+D &           & pair               & bridge,tail      \\      
3C 338   &       & 6166   & cD   & filaments & A2199 CC cluster   & nest             \\      
3C 386   &       &        & E    &           & single             &                  \\      
3C 405   & Cyg A &        & Ep   & complex   & Cyg A cluster      & merger           \\      
3C 424   &       &        & Ep   &           & group              &                  \\      
3C 433   &       &        & S0   & complex   & group              & bridge,nest      \\      
3C 436   &       &        & Ep   & lane      & single             & merger           \\      
Cen A    &       & 5128   & Ep   & lanes     & single             & merger           \\      
\enddata
\tablenotetext{a}{Host morphology. E$=$ elliptical, Ep$=$ peculiar elliptical, cD $=$ cluster dominant, 
                  S0 $=$ lenticular. The tag '+D' indicates a significant exponential disk component in 
                  the near-IR radial light profile \citep{dcm07}.}
\tablenotetext{b}{Dust morphology, after \citep{dbb00}.}
\tablenotetext{c}{Cluster, group, or pair membership. CC cluster $=$ cool X-ray core cluster. }
\tablenotetext{d}{Qualitative assessment of tidal interactions including close companion, tail, bridge, or 
                  merger. 'Nest' indicates multiple galaxies surrounded by a common envelope.}
\end{deluxetable}

\clearpage
\begin{deluxetable}{lcccccccc}
\tablecaption{Radio MOHEG H$_2$ 0-0 S(J) Emission Lines}
\tablehead{
\colhead{Source} &  \colhead{H$_2$ S(0)} &  \colhead{H$_2$ S(1)} &  \colhead{H$_2$ S(2)} &  \colhead{H$_2$ S(3)} &  \colhead{H$_2$ S(4)} &  \colhead{H$_2$ S(5)} &  \colhead{H$_2$ S(6)} &  \colhead{H$_2$ S(7)}\\
             &  28.22 $\mu$m         &  17.03 $\mu$m         &  12.28 $\mu$m         &  9.66 $\mu$m          &  8.03 $\mu$m          &  6.91 $\mu$m          &  6.11 $\mu$m                &  5.55 $\mu$m             }
\startdata
3C 31    &  0.64(0.11) &  1.36(0.58) &  0.59(0.15) &  0.60(0.17) &  0.76(0.27) &  0.80(0.18) &  $<0.9$     &  $<0.4$     \\ 
3C 84    &  $<4.$      & 25. (2.)    &  6.7 (0.5)  & 43. (5.)    &  8.2 (1.1)  & 15. (1.)    &  2.9 (0.9)  &  8.8 (0.9)  \\ 
3C 218   &  0.55(0.11) &  0.41(0.22) &  0.23(0.09) &  0.55(0.08) &  0.42(0.15) &  0.23(0.05) &  \nodata    &  0.16(0.05) \\ 
3C 270   &  $<0.4$     &  0.27(0.11) &  0.57(0.08) &  0.47(0.20) &  $<0.3$     &  0.81(0.26) &  0.37(0.19) &  $<0.7$     \\ 
3C 272.1 &  0.42(0.14) &  0.55(0.29) &  0.27(0.15) &  1.4 (0.2)  &  $<0.6$     &  1.0 (0.4)  &  1.6 (0.4)  &  $<0.6$     \\ 
3C 293   &  1.6 (0.2)  &  5.3 (0.4)  &  1.94(0.07) &  3.25(0.07) &  1.2 (0.1)  &  2.8 (0.2)  &  0.31(0.11) &  1.3 (0.1)  \\ 
3C 310   &  0.12(0.06) &  0.36(0.07) &  0.08(0.04) &  0.48(0.04) &  0.19(0.06) &  $<0.3$     &  $<0.14$    &  0.29(0.09) \\ 
3C 315   &  $<0.12$    &  1.0 (0.1)  &  0.30(0.04) &  0.50(0.04) &  $<0.11$    &  0.28(0.15) &  0.27(0.09) &  $<0.3$     \\ 
3C 317   &  $<0.12$    &  0.44(0.07) &  0.25(0.04) &  0.44(0.05) &  0.25(0.07) &  0.58(0.11) &  0.44(0.11) &  0.17(0.09) \\ 
3C 326 N &  0.30(0.06) &  0.69(0.06) &  0.41(0.04) &  1.26(0.05) &  0.31(0.06) &  0.46(0.22) &  0.25(0.09) &  0.29(0.09) \\ 
3C 338   &  $<0.2$     &  0.34(0.06) &  0.32(0.10) &  0.81(0.12) &  $<0.3$     &  0.20(0.10) &  $<0.2$     &  $<0.2$     \\ 
3C 386   &  $<0.2$     &  0.43(0.06) &  0.09(0.04) &  0.46(0.05) &  $<0.11$    &  0.37(0.10) &  0.30(0.06) &  0.18(0.07) \\ 
3C 405   &  $<1.0$     &  1.5(0.4)   &  $<1.9$     &  4.1 (0.5)  &  $<3.$      &  3.5 (0.9)  &  1.6 (0.7)  &  4.4 (0.8)  \\ 
3C 424   &  0.45(0.07) &  1.12(0.06) &  0.21(0.03) &  0.43(0.04) &  0.15(0.04) &  0.17(0.07) &  $<0.18$    &  0.28(0.05) \\ 
3C 433   &  2.0 (0.4)  &  1.4 (0.2)  &  0.58(0.17) &  1.2 (0.2)  &  $<0.6$     &  1.6 (0.5)  &  $<0.5$     &  1.2 (0.3)  \\ 
3C 436   &  0.48(0.16) &  0.46(0.07) &  0.31(0.12) &  0.28(0.04) &  $<0.10$    &  0.22(0.10) &  0.41(0.12) &  0.21(0.10) \\ 
Cen A    &  10. (6.)   & 57. (9.)    & 27. (4.)    &  9.3(3.7)   &  $<13.$     & 12. (3.)    &  $<8.$      &  27. (4.)   \\ 
\enddata
\tablecomments{H$_2$ pure rotational line fluxes ($10^{-14}$ erg s$^{-1}$ cm$^{-2}$) or $2\sigma$ upper limits  
                  measured with {\it Spitzer} IRS, with $1\sigma$ uncertainties in parentheses. The 3C 218 H$_2$ S(6) line was
                  unmeasurable owing to a large instrumental noise spike.}
\end{deluxetable}

\begin{deluxetable}{lrrrrrrr}
\tablecaption{Radio MOHEG PAH Features}
\tablehead{
 \colhead{Source} & \colhead{6.2 $\mu$m} & \colhead{7.7 $\mu$m} & \colhead{8.6 $\mu$m} & \colhead{11.3 $\mu$m} &  \colhead{12.0 $\mu$m} &  \colhead{12.6  $\mu$m} &  \colhead{17 $\mu$m }}  
\startdata
3C  31  & 22.7(0.5) & 108.(2.)   &  20.0(0.8) & 50.5(0.5) & 16.4(0.7) & 27.3(0.7) & 28.4(1.0) \\
3C  84  & 20. (3.)  & 167.(12.)  &  19. (3.)  & 82. (2.)  & 61. (2.)  & 34. (3.)  & 38. (10.) \\
3C 218  &  3.5(0.2) &  14.(2.)   &   4.0(0.5) &  8.8(0.4) & $<0.8$    &  3.9(0.5) &  2.8(1.3) \\
3C 270  &  5.2(0.6) &  17.(2.)   &  $<0.9$    &  7.0(0.2) &  5.7(0.4) &  4.1(0.4) & $<0.5$    \\
3C 272.1& 16. (1.)  &  21.(4.)   &  $<6.$     & 28.9(0.5) &  5.2(0.6) & 11.4(0.7) & 22.1(1.4) \\
3C 293  & 12.0(0.6) &  50.(2.)   &  21.9(0.6) & 42.7(0.5) & 14.4(0.5) & 23.6(0.6) & 26.2(1.5) \\
3C 310  & 0.56(0.28)&  $<1.4$    &  $<0.3$    & 0.42(0.12)& 0.29(0.13)& 0.15(0.06)& $<0.3$    \\
3C 315  &  1.6(0.4) &   3.6(1.0) &  0.52(0.23)&  3.0(0.2) & 0.87(0.22)&  1.9(0.3) &  4.8(0.6) \\
3C 317  &  $<1.3$   &  $<2.$     &  $<0.8$    & 0.85(0.13)& 0.92(0.20)&  1.3(0.2) &  1.5(0.4) \\
3C 326 N&  $<0.5$   &  $<0.6$    &  $<0.4$    & 0.57(0.12)&  $<0.3$   & $<0.2$    & $<0.7$    \\
3C 338  & 0.70(0.32)&  $<3.$     &  $<1.0$    &  1.9(0.3) &  1.0(0.4) & $<0.8$    & $<0.8$    \\
3C 386  & 0.44(0.21)&  $<2.$     &  $<0.6$    &  1.1(0.1) &  $<0.4$   & 0.72(0.18)& $<0.7$    \\
3C 405  &  $<6.$    &  $<22.$    &  $<6.$     &  $<11.$   &  $<14.$   & $<12.$    & $<14.$    \\  
3C 424  &  $<0.3$   &  $<0.6$    &  $<0.2$    & 0.42(0.08)& 0.57(0.11)& 0.87(0.15)& $<0.6$    \\
3C 433  &  2.5(1.1) &   8. (4.)  &   9. (1.)  &  3.6(0.8) &   9.5(1.0)& 13. (1.)  & 32. (2.)  \\
3C 436  &  $<0.7$   &   3.2(1.6) &  $<0.8$    &  1.1(0.4) & $<1.4$    &  2.1(0.7) & $<1.6$    \\
 Cen A  & 198.(17.) &1052.(92.)  & 459.(32.)  &1557.(26.) & 848.(24.) & 655.(27.) &2668.(73.) \\

\enddata
\tablecomments{Flux ($10^{-14}$ erg s$^{-1}$ cm$^{-2}$) or 2$\sigma$ upper limit measured with {\it Spitzer} IRS and PAHFIT, with $1\sigma$ uncertainties 
in parentheses.}
\end{deluxetable}

\clearpage
\begin{deluxetable}{lccccccc}
\tablecaption{H$_2$ Upper Limits or Single-Line Detections and PAH Features}
\tablehead{
 \colhead{Source} &  \colhead{H$_2$ S(0)} &  \colhead{H$_2$ S(1)} &  \colhead{H$_2$ S(2)} &  \colhead{H$_2$ S(3)} &  \colhead{PAH 6.2 $\mu$m} &  \colhead{PAH 7.7 $\mu$m} &  \colhead{PAH 11.3 $\mu$m} }
\startdata
3C 15    &  $<0.4$     &  $<0.4$     &  $<0.10$    &  $<0.09$    &   $<0.9$    &   $<2.$     &   0.57 (0.13) \\ 
3C 29    &  0.55(0.15) &  $<0.7$     &  $<0.2$     &  $<0.4$     &   $<0.9$    &   $<3.$     &   $<0.44$     \\ 
3C 66B   &  $<0.12$    &  0.30(0.07) &  $<0.10$    &  $<0.11$    &   $<0.5$    &   $<0.7$    &   0.73 (0.10) \\ 
3C 76.1  &  $<0.2$     &  0.45(0.09) &  $<0.12$    &  $<0.06$    &   $<1.3$    &   $<3.$     &   1.48 (0.11) \\ 
3C 83.1  &  $<0.14$    &  $<0.14$    &  $<0.11$    &  0.25(0.06) &   1.5 (0.5) &   3.2 (1.6) &   2.7  (0.2)  \\ 
3C 120   &  $<2.2$     &  2.74(0.75) &  $<1.5$     &  $<1.0$     &   7.9 (2.0) & 117.  (14.) &  33.   (2.)   \\ 
3C 129   &  $<0.3$     &  $<0.19$    &  $<0.16$    &  0.31(0.08) &   $<0.9$    &   $<1.5$    &   2.9  (0.2)  \\ 
3C 189   &  0.23(0.08) &  $<0.4$     &  $<0.11$    &  $<0.3$     &   $<1.3$    &   $<1.7$    &   $<0.26$     \\ 
3C 264   &  $<0.3$     &  0.28(0.12)\tablenotemark{a} &  $<0.11$    &  0.20(0.05) &   7.6 (0.5) &  14.9 (1.7) &   6.2  (0.2)  \\ 
3C 274   &  $<0.7$     &  $<1.1$     &  $<1.6$     &  2.92(0.62) &   $<7.$     &   $<28.$    &   $<3.$       \\ 
3C 318.1 &  $<0.14$    &  $<0.16$    &  $<0.06$    &  $<0.06$    &   $<0.7$    &   $<1.2$    &   0.31 (0.07) \\ 
3C 348   &  $<0.4$     &  $<0.5$     &  $<0.4$     &  $<0.5$     &   $<2.$     &   $<3.$     &   $<1.6$      \\ 
3C 403.1 &  $<0.2$     &  $<0.16$    &  $<0.08$    &  0.10(0.03) &   $<0.4$    &   $<1.1$    &   0.65 (0.10) \\ 
3C 449   &  $<0.09$    &  $<0.4$     &  $<0.08$    &  $<0.04$    &   $<0.2$    &   $<0.6$    &   0.24 (0.07) \\
3C 465   &  $<0.18$    &  $<0.2$     &  $<0.07$    &  $<0.10$    &   $<0.4$    &   $<1.1$    &   0.58 (0.10) \\ 
NGC 6251 &  $<0.3$     &  $<0.3$     &  $<0.18$    &  $<0.2$     &   1.6 (0.2) &   6.6 (1.0) &   2.7  (0.2)  \\ 
IC 4296  &  $<0.3$     &  $<0.3$     &  $<0.2$     &  0.78(0.15) &   $<3.$     &   5.6 (1.0) &   2.6  (0.3)  \\ 
3C 17    &  $<0.4$     &  $<0.5$     &  $<0.5$     &  $<0.2$     &   $<1.2$    &   $<2.$     &   1.6  (0.3)  \\ 
3C 28    &  $<0.3$     &  $<0.4$     &  $<0.4$     &  0.22(0.04) &   $<1.8$    &   $<1.2$    &   5.8  (0.6)  \\ 
3C 33    &  $<0.7$     &  $<0.5$     &  $<0.19$    &  0.25(0.09) &   $<2.$     &   $<6.$     &   5.2  (0.4)  \\ 
3C 61.1  &  $<0.3$     &  0.37(0.10) &  $<0.12$    &  $<0.11$    &   $<1.4$    &   $<3.$     &   $<0.5$      \\ 
3C 111   &  $<1.2$     &  $<1.4$     &  $<1.4$     &  $<0.9$     &   $<6.$     &   $<28.$    &   $<4.$       \\
3C 123   &  $<0.4$     &  $<1.7$     &  $<0.7$     &  $<0.3$     &   $<1.7$    &   $<0.9$    &   $<1.3$      \\ 
3C 192   &  $<0.2$     &  $<0.3$     &  $<0.16$    &  $<0.13$    &   $<1.1$    &   $<1.7$    &   $<0.5$      \\
3C 219   &  $<0.3$     &  0.45(0.09) &  $<0.3$     &  $<0.16$    &   1.9 (0.6) &   $<3.$     &   $<0.5$      \\ 
3C 234   &  $<1.9$     &  \nodata\tablenotemark{b}    &  $<0.9$     &  $<0.4$     &   4.6 (1.6) &   9.2 (4.4) &   $<2.$       \\ 
3C 319   &  $<0.3$     &  $<0.2$     &  $<0.3$     &  $<0.12$    &   $<0.7$    &   $<1.0$    &   0.31 (0.12) \\
3C 321   &  $<1.8$     &  $<1.5$     &  $<1.1$     &  1.44(0.39) &   $<7.$     &  61.  (11.) &   6.   (2.)   \\ 
3C 346   &  $<0.18$    &  $<0.3$     &  $<0.10$    &  $<0.15$    &   $<0.4$    &   $<1.1$    &   $<0.3$      \\ 
3C 381   &  $<1.5$     &  $<1.2$     &  $<0.3$     &  $<0.2$     &   $<1.0$    &   3.8 (1.0) &   1.2  (0.2)  \\
3C 382   &  $<0.4$     &  $<0.6$     &  $<0.4$     &  $<0.7$     &   $<2.$     &   $<6.$     &   $<2.$       \\ 
3C 388   &  $<0.2$     &  $<0.3$     &  $<0.17$    &  $<0.2$     &   $<0.9$    &   $<1.1$    &   $<0.6$      \\
3C 390.3 &  $<0.9$     &  $<0.6$     &  $<0.3$     &  0.24(0.09) &   $<1.8$    &  14.1 (1.5) &   $<2.$       \\ 
3C 401   &  $<0.4$     &  $<0.2$     &  $<0.11$    &  $<0.13$    &   $<0.7$    &   $<1.1$    &   $<0.4$      \\ 
3C 445   &  $<1.8$     &  $<1.9$     &  $<1.6$     &  $<0.8$     &   $<4.$     &   7.5 (3.6) &   $<3.$       \\ 
3C 452   &  $<0.5$     &  $<0.3$     &  $<0.3$     &  $<0.18$    &   $<1.5$    &   3.2 (1.7) &   $<1.2$      \\
3C 459   &  $<2.6$     &  $<1.1$     &  0.94(0.34) &  $<0.4$     &   9.6 (2.1) &  15.0 (1.3) &   9.6  (0.6)  \\ 
Pic A    &  $<0.3$     &  $<0.7$     &  $<0.7$     &  $<0.7$     &   $<6.$     &  $<8.$      &   $<1.6$      \\ 
\enddata
\tablecomments{H$_2$ pure rotational line and PAH fluxes ($10^{-14}$ erg s$^{-1}$ cm$^{-2}$) or $2\sigma$ upper limits measured with {\it Spitzer} IRS, with $1\sigma$ uncertainties in parentheses.}
\tablenotetext{a}{The 3C 264 H$_2$ S(1) line is only a marginal detection (2.3$\sigma$), so this source does not meet our H$_2$ detection criterion of
                  2 lines detected at $>2.5\sigma$.}
\tablenotetext{b}{The 3C 234 H$_2$ S(1) line was unmeasurable owing to a large instrumental noise spike.}
\end{deluxetable}

\clearpage
\begin{turnpage}
\begin{deluxetable}{lccccccccccc}
\tablecaption{Radio MOHEG Forbidden Emission Lines}
\tablehead{
\colhead{Source} & \colhead{[Fe {\sc ii}]} & \colhead{[Ar {\sc ii}]} & \colhead{[Ar {\sc iii}]} & \colhead{[S {\sc iv}]} &  \colhead{[Ne {\sc ii}]} &  \colhead{[Cl {\sc ii}]} &  \colhead{[Ne {\sc iii}]} &  \colhead{[S {\sc iii}]} & \colhead{[O {\sc iv}]} & \colhead{[S {\sc iii}]} &  \colhead{[Si {\sc ii}]}\\
\colhead{      } &  \colhead{5.34 $\mu$m  }  &\colhead{6.99 $\mu$m  } & \colhead{8.99 $\mu$m   } & \colhead{10.51 $\mu$m} &  \colhead{12.81 $\mu$m } &  \colhead{14.37 $\mu$m } &  \colhead{15.56 $\mu$m  } &  \colhead{18.71 $\mu$m } & \colhead{25.89 $\mu$m} & \colhead{33.48 $\mu$m } &  \colhead{34.82 $\mu$m }}
\startdata
3C 31   &  0.48(0.18) &  1.3 (0.2)  &    $<0.3$   &  0.51(0.18) &  3.0 (0.2)  &    $<0.4$   &  1.5 (0.2)  &  0.73(0.17) &  1.1 (0.2) &  0.84(0.15) &  2.4 (0.3)   \\
3C 84   & 10.  (1.)   & 12.4(0.9)   &  2.5 (0.8)  &  1.4 (0.6)  & 43.3 (0.5)  &  6.8 (1.7)  & 26.  (2.)   & 13.  (2.)   &    $<7.$    & 23.  (2.)   & 62.3 (0.6)  \\
3C 218  &  0.14(0.05) &  0.46(0.07) &    $<0.3$   &    $<0.2$   &  1.78(0.08) &    $<0.5$   &  0.78(0.17) &    $<0.5$   &  1.0 (0.2)  &  0.93(0.26) &  \nodata    \\
3C 270  &    $<0.8$   &  2.4 (0.3)  &  0.23(0.10) &  0.33(0.08) &  4.20(0.08) &  0.58(0.20) &  2.3 (0.1)  &  0.87(0.11) &  1.3 (0.3)  &  1.3 (0.2)  &  2.9 (0.1)  \\
3C 272.1&    $<1.5$   &  3.8 (0.6)  &    $<0.6$   &  1.0 (0.2)  &  5.6 (0.2)  &  0.44(0.20) &  5.7 (0.2)  &  3.4 (0.2)  &  1.4  (0.2) &  4.1 (0.1)  & 11.7 (0.3)  \\
3C 293  &  1.3 (0.2)  &  1.9 (0.2)  &  0.61(0.09) &    $<0.3$   &  5.24(0.08) &    $<0.3$   &  2.6 (0.1)  &  1.2 (0.2)  &  1.5 (0.1)  &  2.8 (0.3)  &  \nodata    \\
3C 310  &    $<0.2$   &   $<0.3$    &  0.09(0.04) &    $<0.08$  &  0.28(0.04) &    $<0.2$   &  0.47(0.06) &    $<0.2$   &    $<0.2$   &  1.1 (0.2)  &  \nodata    \\
3C 315  &  0.28(0.11) &  0.53(0.14) &  0.12(0.04) &  0.10(0.04) &  0.29(0.05) &    $<0.2$   &  0.30(0.09) &    $<0.3$   &  0.24(0.08) &  \nodata    &  \nodata    \\
3C 317  &    $<0.3$   &  0.65(0.13) &  0.21(0.05) &    $<0.08$  &  0.88(0.05) &  0.19(0.07) &  0.43(0.07) &  0.37(0.07) &  0.26(0.08) &    $<0.3$   &  1.3 (0.2)  \\
3C 326 N&  0.23(0.08) &   $<0.3$    &  0.30(0.05) &  0.09(0.04) &  0.44(0.05) &  0.17(0.05) &    $<0.2$   &    $<0.2$   &  0.12(0.05) &  \nodata    &  \nodata    \\
3C 338  &    $<0.3$   &   $<0.4$    &    $<0.2$   &  0.22(0.11) &  0.66(0.11) &    $<0.2$   &  0.27(0.07) &    $<0.3$   &  0.34(0.10) &  0.58(0.16) &  1.2 (0.3)  \\
3C 386  &  0.43(0.08) &   $<0.2$    &    $<0.1$   &    $<0.1$   &  0.39(0.04) &  0.13(0.06) &  0.61(0.05) &  0.39(0.10) &    $<0.3$   &  0.96(0.21) &  2.9 (0.2)  \\
3C 405  &  2.3 (0.9)  & 11.6 (0.8)  &  4.4 (0.7)  & 16.2 (0.7)  & 21.7 (0.7)  &  5.4 (0.4)  & 47.9 (0.5)  & 24.2 (0.6)  & 78.5 (0.7)  & 29.  (1.)   &  \nodata    \\
3C 424  &  0.15(0.05) &   $<0.07$   &    $<0.09$  &    $<0.07$  &  0.25(0.08) &  0.12(0.06) &  0.16(0.05) &    $<0.2$   &  0.18(0.06) &  \nodata    &  \nodata    \\
3C 433  &    $<0.6$   &  1.3 (0.4)  &  1.5 (0.2)  &  1.7 (0.2)  &  2.6 (0.2)  &    $<0.9$   &  5.7 (0.2)  &  0.89(0.20) &  7.0 (0.4)  &  \nodata    &  \nodata    \\
3C  436 &    $<0.2$   &   $<0.1$    &  0.12(0.05) &  0.12(0.04) &  0.29(0.10) &    $<0.3$   &  0.85(0.10) &  0.28(0.11) &  0.93(0.12) &  \nodata    &  \nodata    \\
Cen A   &    $<18.$   & 67. (4.)    & 10.9 (5.2)  &  28.(4.)    &213.(4.)     & 37. (12.)   &233.(18.)    & 87.  (9.)   &155. (18.)   & 264.(11.)   &  561.(14.)  \\
\enddata
\tablecomments{Flux ($10^{-14}$ erg s$^{-1}$ cm$^{-2}$) or 2$\sigma$ upper limit measured with {\it Spitzer} IRS, with $1\sigma$ uncertainties in parentheses.}
\end{deluxetable}
\end{turnpage}

\begin{deluxetable}{lccc}
\tablecaption{High-Ionization Forbidden Emission Lines}
\tablehead{\colhead{Source} &  \colhead{ [Ne {\sc vi}] 7.65} &  \colhead{[Ne {\sc v}] 14.3} &  \colhead{[Ne {\sc v}] 24.31}}
\startdata

3C 405  & 13. (1.)  & 20.5(0.5) & 24.3(0.6) \\
3C 433  &  2.3(0.4) &  0.8(0.2) &  0.9(0.3) \\  
3C 436  &0.23(0.05) &  $<0.3$   &  $<0.2$   \\
Cen A   & 29.(14.)  & 37.(11.)  & 65.(31.)  \\

\enddata
\tablecomments{Flux ($10^{-14}$ erg s$^{-1}$ cm$^{-2}$) measured with {\it Spitzer} IRS, with $1\sigma$ uncertainties in parentheses.}
\end{deluxetable}

\begin{deluxetable}{lcccc}
\tablecaption{Radio MOHEG Continuum and Line Luminosities}
\tablehead{
\colhead{Source} & \colhead{$\log \nu L_\nu$(178 MHz)}& \colhead{$\log \nu L_\nu$(24 $\mu$m,$L_\odot$)} & \colhead{$\log L$(H$_2$)\tablenotemark{a} } & \colhead{$\log L_\mathrm{X}$(2-10 keV)\tablenotemark{b}}}
\startdata
3C 31    &  40.33 &  8.63  &  40.32 &  40.67 \\
3C 84    &  40.92 & 10.84  &  41.75 &  42.91 \\
3C 218   &  42.46 &  9.34  &  41.10 &  41.69 \\
3C 270   &  40.07 &  8.23  &  39.26 &  41.08 \\
3C 272.1 &  39.02 &  7.42  &  38.86 &  39.34 \\
3C 293   &  41.07 &  9.74  &  41.76 &  42.78 \\
3C 310   &  41.87 &  8.64  &  40.85 & \nodata \\
3C 315   &  42.02 &  9.43  &  41.76 &  41.68\tablenotemark{c} \\
3C 317   &  41.42 &  8.61  &  40.53 &  41.30 \\
3C 326 N &  41.90 &  8.57  &  41.73 &  40.63\tablenotemark{c} \\
3C 338   &  41.28 &  8.49  &  40.54 &  40.30 \\
3C 386   &  40.48 &  7.75  &  39.87 &  39.75\tablenotemark{c}\\
3C 405   &  44.10 & 11.31  &  41.75 &  44.28 \\
3C 424   &  42.08 &  9.53  &  41.97 & \nodata \\
3C 433   &  42.46 & 11.15  &  42.13 &  43.90\tablenotemark{c}\\
3C 436   &  42.67 & 10.32  &  42.31 &  43.53 \\
Cen A    &  40.12 &  9.07  &  39.88 &  41.70 \\
\enddata
\tablenotetext{a}{Observed H$_2$ rotational line luminosity (erg s$^{-1}$), summed over the 0-0 S(0)-S(3) transitions (Table 4). Corrections of 6-33\% are made for 
                  undetected emission lines, based on H$_2$ multi-temperature models (\S 5).}
\tablenotetext{b}{Unabsorbed 2-10 keV nuclear X-ray luminosities (erg s$^{-1}$) from \cite{ewh06}, \cite{bcg06}, \cite{hec06}, \cite{mwn00}, \cite{pell05}, and \cite{uit05} }  
\tablenotetext{c}{Unabsorbed 2-10 keV nuclear X-ray luminosities (erg s$^{-1}$)  measured with {\it Chandra} (Table 10).}                                                               
\end{deluxetable}

\begin{deluxetable}{llllll}
\tablecaption{Chandra X-ray Data and AGN Spectral Models}
\tablehead{\colhead{Source} & \colhead{$t$ (ks)} & \colhead{chip} & \colhead{model}  & 
\colhead{$L_\mathrm{X}$\tablenotemark{a}} & \colhead{Notes}}
\startdata
3C 315  & 7.7       & S3          & PL       & $4.8\times10^{41}$ & S/N too
low to detect heavily abs. emission \\
3C 326 N& 27.5      & I3          & PL       & $4.3\times10^{40}$ & Low S/N \\
3C 386  & 29.1      & I3          & PL       & $5.6\times10^{39}$ & \\
3C 433  & 37.2      & S3          & PL + abs. PL &
$8.0\times10^{43}$ & See Miller \& Brandt (2009) \\
Zw 3146 & 45.6     & I3          & PL       & $8.2\times10^{42}$ & Likely
contamination from thermal gas\\
\enddata
\tablenotetext{a}{Unabsorbed 2-10 keV nuclear X-ray luminosities (erg s$^{-1}$), measured with {\it Chandra}.}              
\end{deluxetable}

\clearpage
\begin{deluxetable}{lrccrc}
\tablecaption{H$_2$ Model Parameters}
\tablehead{
\colhead{Source} & \colhead{T(K)\tablenotemark{a} }& \colhead{Ortho/Para\tablenotemark{b}}& 
\colhead{$N($H$_2)$\tablenotemark{c}} & \colhead{$M($H$_2)$\tablenotemark{d}}& \colhead{$L_\mathrm{tot}($H$_2,\mathrm{rot})$\tablenotemark{e}}}
\startdata
 
3C 31    &  100.  (  0.) & 1.587 & 8.6E+21 & 2.3E+08 (1.6E+08,3.1E+08) & 5.8E+39 \\
         &  250.  ( 60.) & 2.955 & 2.3E+20 & 6.0E+06 (1.8E+06,2.0E+07) & 1.6E+40 \\
         & 1140.  (150.) & 3.000 & 9.5E+17 & 2.5E+04 (1.5E+04,4.0E+04) & 2.3E+40 \\ 
 
3C 84    &  150.  ( 20.) & 2.464 &$<$3E+22 &$<$8E+08                   & 2.0E+41 \\ 
         &  580.  ( 30.) & 3.000 & 1.7E+20 & 4.7E+06 (4.0E+06,5.6E+06) & 3.8E+41 \\ 
         & 1500.  (  0.) & 3.000 & 3.2E+18 & 9.0E+04 (8.0E+04,1.0E+05) & 2.2E+41 \\ 
 
3C 218   &  100.  ( 10.) & 1.654 & 8.2E+21 & 2.0E+09 (1.2E+09,3.5E+09) & 7.2E+40 \\ 
         &  540.  ( 70.) & 3.000 & 4.7E+18 & 1.2E+06 (7.0E+05,2.0E+06) & 8.4E+40 \\ 
         & 1500.  (  0.) & 3.000 & 5.9E+16 & 1.5E+04 (1.0E+04,2.1E+04) & 4.0E+40 \\ 
 
3C 270   &  100.  (  0.) & 1.587 &$<$5E+21 &$<$3E+07                   & 6.7E+38 \\ 
         &  400.  ( 70.) & 2.999 & 3.1E+19 & 1.6E+05 (8.8E+04,2.8E+05) & 2.8E+39 \\ 
         & 1500.  (  0.) & 3.000 & 3.3E+17 & 1.7E+03 (1.4E+03,2.1E+03) & 3.9E+39 \\ 
 
3C 272.1 &  100.  (  0.) & 1.587 & 6.7E+21 & 7.9E+06 (5.3E+06,1.2E+07) & 1.9E+38 \\ 
         &  290.  (180.) & 2.983 & 2.7E+19 & 3.2E+04 (3.4E+03,3.0E+05) & 1.5E+38 \\ 
         & 1150.  (120.) & 3.000 & 1.9E+18 & 2.2E+03 (1.5E+03,3.2E+03) & 2.0E+39 \\ 
 
3C 293   &  100.  (  0.) & 1.587 & 2.1E+22 & 3.6E+09 (3.1E+09,4.2E+09) & 1.0E+41 \\ 
         &  278.  (  9.) & 2.980 & 4.8E+20 & 8.3E+07 (7.0E+07,9.7E+07) & 4.1E+41 \\ 
         & 1040.  ( 30.) & 3.000 & 3.2E+18 & 5.4E+05 (4.8E+05,6.2E+05) & 4.1E+41 \\ 
 
3C 310   &  140.  ( 20.) & 2.340 & 7.1E+20 & 1.7E+08 (9.0E+07,3.2E+08) & 3.4E+40 \\ 
         &  640.  (160.) & 3.000 & 1.7E+18 & 4.1E+05 (2.1E+05,8.0E+05) & 5.7E+40 \\ 
         & 1500.  (  0.) & 3.000 & 1.1E+17 & 2.7E+04 (1.9E+04,3.8E+04) & 7.4E+40 \\ 
 
3C 315   &  160.  ( 40.) & 2.603 &$<$4E+20 &$<$3E+08                   & 1.8E+41 \\ 
         &  330.  ( 40.) & 2.994 & 3.6E+19 & 3.1E+07 (1.4E+07,7.0E+07) & 3.8E+41 \\ 
         & 1500.  (  0.) & 3.000 & 1.6E+17 & 1.4E+05 (1.1E+05,1.7E+05) & 4.6E+41 \\ 
 
3C 317   &  100.  (  0.) & 1.587 &$<$2E+21 &$<$2E+08                   & 4.3E+39 \\
         &  300.  ( 40.) & 2.988 & 3.4E+19 & 3.5E+06 (2.2E+06,5.5E+06) & 2.2E+40 \\
         & 1220.  (150.) & 3.000 & 4.3E+17 & 4.4E+04 (3.0E+04,6.7E+04) & 5.7E+40 \\
 
3C 326 N &  110.  ( 10.) & 1.751 & 3.7E+21 & 2.2E+09 (1.3E+09,3.8E+09) & 1.1E+41 \\
         &  430.  ( 60.) & 3.000 & 1.9E+19 & 1.2E+07 (7.7E+06,1.7E+07) & 3.9E+41 \\
         & 1110.  (250.) & 3.000 & 5.6E+17 & 3.4E+05 (1.2E+05,9.7E+05) & 3.8E+41 \\
 
3C 338   &  100.  (  0.) & 1.587 &$<$3E+21 &$<$2E+08                   & 6.6E+39 \\
         &  480.  ( 60.) & 3.000 & 9.4E+18 & 7.6E+05 (5.1E+05,1.1E+06) & 3.3E+40 \\
         & 1450.  (590.) & 3.000 &$<$1E+17 &$<$1E+04                   & 2.2E+40 \\
 
3C 386   &  130.  ( 20.) & 2.124 &$<$2E+21 &$<$4E+07                   & 3.9E+39 \\
         &  430.  (150.) & 3.000 & 5.2E+18 & 1.4E+05 (2.8E+04,6.4E+05) & 3.6E+39 \\
         & 1500.  (  0.) & 3.000 & 2.1E+17 & 5.5E+03 (4.6E+03,6.5E+03) & 1.3E+40 \\
 
3C 405   &  100.  (  0.) & 1.587 &$<2$E+22 &$<$4E+09                   & 1.2E+41 \\
         &  460.  ( 70.) & 3.000 & 4.5E+19 & 1.2E+07 (6.8E+06,2.1E+07) & 4.6E+41 \\
         & 1500.  (  0.) & 3.000 & 1.8E+18 & 4.6E+05 (3.9E+05,5.3E+05) & 1.3E+42 \\
 
3C 424   &  132.  (  8.) & 2.237 & 2.9E+21 & 3.3E+09 (2.4E+09,4.4E+09) & 6.5E+41 \\
         &  360.  ( 60.) & 2.997 & 1.6E+19 & 1.8E+07 (8.5E+06,4.0E+07) & 3.4E+41 \\
         & 1340.  (230.) & 3.000 & 2.0E+17 & 2.2E+05 (1.2E+05,4.3E+05) & 5.4E+41 \\
 
3C 433   &  100.  (  0.) & 1.587 & 3.1E+22 & 2.3E+10 (1.9E+10,2.8E+10) & 8.3E+41 \\
         &  350.  ( 80.) & 2.996 & 4.2E+19 & 3.2E+07 (1.4E+07,7.4E+07) & 4.7E+41 \\
         & 1300.  (210.) & 3.000 & 9.3E+17 & 7.1E+05 (3.9E+05,1.3E+06) & 1.4E+42 \\
 
3C 436   &  100.  (  0.) & 1.587 & 6.2E+21 & 1.6E+10 (1.1E+10,2.4E+10) & 8.6E+41 \\
         &  300.  ( 40.) & 2.987 & 2.9E+19 & 7.6E+07 (4.1E+07,1.4E+08) & 8.6E+41 \\
         & 1500.  (  0.) & 3.000 & 2.0E+17 & 5.2E+05 (4.3E+05,6.4E+05) & 2.5E+42 \\
 
Cen A    &  100.  (  0.) & 1.587 & 1.1E+23 & 3.4E+07 (1.9E+07,5.9E+07) & 8.2E+38 \\
         &  290.  ( 30.) & 2.986 & 5.0E+21 & 1.6E+06 (1.1E+06,2.3E+06) & 8.0E+39 \\
         & 1500.  (  0.) & 3.000 & 9.6E+18 & 3.0E+03 (2.7E+03,3.4E+03) & 6.7E+39 \\
 
\enddata
\tablenotetext{a}{H$_2$ model component temperatures. A temperature range of $100<T<1500$ K was enforced in fitting the models.
Temperatures pegged at either end of this range have indeterminate uncertainties.}                                     
\tablenotetext{b}{Ratio of ortho (odd J) to para (even J) H$_2$, fit self-consistently to match the thermal equilibrium value.}                
\tablenotetext{c}{H$_2$ model column densities (cm$^{-2}$), assuming a source size of $3\farcs7 \times  3\farcs7$, matching the SL slit width.} 
\tablenotetext{d}{H$_2$ model masses and 1$\sigma$ mass ranges ($M_\odot$).} 
\tablenotetext{e}{H$_2$ model luminosities (erg s$^{-1}$), summed over all pure-rotational transitions. Note that not all of these transitions
                  are directly observed by {\it Spitzer}.} 
\end{deluxetable}

\clearpage
\begin{deluxetable}{lcccccc}
\tablecaption{Radio MOHEG H$_2$ and Dust Masses}
\tablehead{
\colhead{Source} & \colhead{$M($warm H$_2)$\tablenotemark{a}} &\colhead{$M($cold H$_2)$\tablenotemark{b}} &\colhead{warm/cold\tablenotemark{c}} &\colhead{Ref. \tablenotemark{d}} &\colhead{$M_\mathrm{dust}$\tablenotemark{e}}
}

\startdata
3C 31    & 2.4E+08  &  1.3E+09  & 0.18      & 4 & 2.0E+05 \\                 
3C 84    & 4.8E+06-8E+08 &  2.2E+09  &$<0.36$    & 5 & 4.0E+05 \\                 
3C 218   & 2.0E+09  &  2.0E+09  & 1.0       & 2 & \nodata \\                 
3C 270   & 1.6E+05-3E+07 & $<$9E+07  & \nodata   & 8 & 1.0E+03 \\                 
3C 272.1 & 7.9E+06  &  3.7E+06: & 2.1:      & 8 & 4.0E+03 \\                 
3C 293   & 3.7E+09  &  2.3E+10  & 0.16      & 1 & 2.0E+06 \\                 
3C 310   & 1.7E+08  &  \nodata  & \nodata   & \nodata & \nodata \\           
3C 315   & 3.1E+07-3E+08 &  \nodata  & \nodata   & \nodata & 1.3E+05 \\           
3C 317   & 3.5E+06-2E+08 &  \nodata  & \nodata   & \nodata & \nodata \\           
3C 326 N & 2.2E+09  &  1.7E+09  & 1.3       & 9 & \nodata \\                 
3C 338   & 7.6E+05-2E+08 & $<$1E+09  & \nodata   & 2 & 1.3E+05 \\                 
3C 386   & 1.5E+05-4E+07 &  2.2E+08  &$<0.18$    & 8 & \nodata \\                 
3C 405   & 1.2E+07-4E+09 & $<$1E+09  & \nodata   & 3 & \nodata \\                 
3C 424   & 3.3E+09  & $<6$E+09  & $>$0.6    & 6 & \nodata \\                 
3C 433   & 2.3E+10  & $<9$E+09  & $>$3      & 3 & 5.0E+05 \\                 
3C 436   & 1.6E+10  & \nodata   & \nodata   & \nodata & \nodata \\           
Cen A    & 3.6E+07  & $<$3E+08\tablenotemark{f}& $>$0.12  & 7 & 5.0E+05\\  
\enddata
\tablenotetext{a}{Total mass ($M_\odot$) of warm H$_2$ components from Table 11. Mass ranges are given for galaxies that are undetected in H$_2$ S(0)
                  but detected in other H$_2$ rotational lines, since this line determines our estimate of the mass of the coldest and typically most
                  massive warm H$_2$ component.}
\tablenotetext{b}{Cold H$_2$ mass ($M_\odot$), estimated from CO flux density and corrected to the same standard Galactic CO conversion factor of 
4.6 $M_\odot/$(K km s$^{-1}$ pc$^2$). Upper limits may be underestimated in the case of broad CO lines. Tentative (low S/N) detections are 
indicated by a colon.}
\tablenotetext{c}{Ratio of cols. 2 and 3.}
\tablenotetext{d}{Cold H$_2$ Refs: 1) \cite{ess99} 2) \cite{sc03} 3) \cite{ems05} 4) \cite{oki05} 5) \cite{sce06} 6) \cite{sm07},
7) \cite{ivb90} 8) \cite{olc10} 9) \cite{nbs10}.}
\tablenotetext{e}{Mass ($M_\odot$) of clumpy dust estimated from optical absorption \citep{dbb00}. This could be much less than the amount of dust
associated with the warm H$_2$ or CO, which might be revealed with sensitive far-IR continuum measurements.} 
\tablenotetext{f}{Cen A cold H$_2$ mass detected within 40'' aperture.}                                     
\end{deluxetable}

\clearpage
\begin{deluxetable}{lll}
\tablecaption{Star Formation Rates}
\tablehead{
\colhead{Source}&  \colhead{SFR(PAH 7.7)} &  \colhead{SFR(PAH 11.3)} 
}

\startdata
3C  31   & 0.43   (0.01)   & 0.78  (0.01)   \\
3C  84   & 0.71   (0.05)   & 1.35  (0.03)   \\
3C 218   & 0.62   (0.09)   & 1.49  (0.07)   \\
3C 270   & 0.013  (0.002)  & 0.021 (0.001)  \\
3C 272.1 & 0.0036 (0.0007) & 0.019 (0.001)  \\
3C 293   & 1.47   (0.06)   & 4.81  (0.06)   \\
3C 310   & $<0.06$         & 0.068 (0.020)  \\
3C 315   & 0.67   (0.19)   & 2.1   (0.1)    \\
3C 317   & $<0.03$         & 0.055 (0.009)  \\
3C 326 N & $<0.07$         & 0.27  (0.06)   \\
3C 338   & $<0.04$         & 0.10  (0.02)   \\
3C 386   & $<0.008$        & 0.017 (0.002)  \\
3C 405   & $<1.  $         & $<1.9$         \\
3C 424   & $<0.16$         & 0.42  (0.08)   \\
3C 433   & 1.3    (0.6)    & 2.2   (0.5)    \\
3C 436   & 2.6    (1.3)    & 3.5   (1.3)    \\
Cen A    & 0.048  (0.004)  & 0.27  (0.01)   \\
\enddata
\tablecomments{Star formation rates ($M_\odot~\mathrm{yr}^{-1}$) are estimated from 
                  SFR(PAH 7.7)$=2.4\times10^{-9} L(\mathrm{PAH7.7})/L_\odot$ and
                  SFR(PAH 11.3)$=9.2\times10^{-9} L(\mathrm{PAH11.3})/L_\odot$, using
                  PAH luminosities inside the {\it Spitzer} IRS SL slit. SFR(PAH 11.3) should be considered
                  an upper limit for sources with high ($>0.26$) 11.3/7.7 PAH ratio.}
\end{deluxetable}

\begin{deluxetable}{llcc}
\tablecaption{Jet Cavity Power}
\tablehead{
\colhead{Source} &  \colhead{Cluster} &  \colhead{P(jet cavity)\tablenotemark{a}} & \colhead{L(H$_2$)/P(jet cavity)\tablenotemark{b}} \\ 
}

\startdata
3C 84    & Perseus A   & 1.4E+44  & 2.7E-03 \\
3C 218   & Hydra A     & 4.3E+44  & 3.4E-04 \\
3C 272.1 & M84, Virgo  & 1.0E+42  & 1.1E-03 \\
3C 317   & A2052       & 1.5E+44  & 2.3E-04 \\ 
3C 338   & A2199       & 2.7E+44  & 1.2E-04 \\
3C 405   & Cyg A       & 1.3E+45  & 3.8E-04\\
\enddata
\tablenotetext{a}{Jet cavity power estimated from the $4pV$ enthalpy required to inflate the cavity, divided by the cavity buoyancy timescale 
                  \citep{rmn06}.}
\tablenotetext{b}{Ratio of H$_2$ pure rotational line luminosity (summed over 0-0 S(0)-S(3)) to jet cavity power.}
\end{deluxetable}


\end{document}